\documentstyle[prb,aps,epsfig,eqsecnum]{revtex}

\begin{document}

\title{Shot Noise and the Transmission of Dilute Laughlin Quasiparticles}

\author{C.L. Kane} \address{Department of Physics and Astronomy,
University of Pennsylvania, Philadelphia, PA 19104}

\author{Matthew P.A. Fisher}
\address{Institute for Theoretical Physics,
University of California, Santa Barbara, CA 91036}

\date{August 30, 2002}
\maketitle

\draft

\begin{abstract}
We analyze theoretically a three-terminal geometry in a fractional quantum Hall
system - studied in a recent experiment - which allows a dilute beam of Laughlin
quasiparticles to be prepared and subsequently scattered by a point contact.  Employing
a chiral Luttinger liquid description of the $\nu^{-1} =m$ integer edge states, we
compute the current and noise of the quasiparticle beam after transmission through
the point contact at finite temperature and bias voltage.  A re-fermionization procedure at
$m=2$ allows the current and noise to be computed non-perturbatively for arbitrary
transparency of the point contact.  Surprisingly, we find for weak backscattering
the zero temperature limit is subtle and singular even at fixed finite bias voltage.
In particular, at $T=0$ the incident charge $e/m$ quasiparticles are either reflected
or else {\it Andreev} scattered (backscattering a charge $(-1 +1/m)e$ quasihole and transmitting
an electron) -  Laughlin quasiparticles are {\it not} transmitted in this limit.
A direct signature of these Andreev processes should be accessible in a particular
cross-correlation noise measurement that we propose.

\end{abstract}


\section{Introduction}

One of the most striking consequences of strong correlation in electronic systems is
charge fractionalization, where the elementary charged excitations of a system have quantum
numbers which differ from those of the bare electron.  The fractional quantum Hall effect is an ideal
arena to study this phenomena\cite{review}.  At filling factor $\nu = 1/m$, the elementary excitation of
the quantum Hall state is the charge $e/m$ Laughlin quasiparticle\cite{laughlin}.  Current experimental
techniques allow for a detailed study of the transport properties of these exotic particles.

A powerful technique for probing elementary charge carriers is to measure shot noise.
When particles flow independently with an uncorrelated Poisson distribution, their charge
is given by the ratio between the mean square fluctuation of the current
and the average current\cite{shottky}.  In 1994 we proposed that a quantum point contact, formed by pinching
together the edges of a quantum Hall bar, would be an ideal geometry for establishing the
uncorrelated flow of Laughlin  quasiparticles\cite{kfnoise}.  When the point contact is strongly pinched
off the sample is effectively split into two.  In that case a weak tunneling current must be
carried by electrons, and shot noise with charge $e$ is expected.
However, in the opposite extreme of weak pinch off, quasiparticles can backscatter between the edges
through the quantum Hall fluid.  The ratio between the noise and the backscattered current is then
determined by the charge of the quasiparticle.  In seminal 1997 experiments, de-Piccioto et al.
\cite{depiccioto} and Saminadayar, et al. \cite{saminadayar}
 independently used this technique to measure the charge $e/3$ of the
Laughlin quasiparticle.

The original experiments used a two terminal setup in which the current and noise transmitted
through the point contact were measured.  The backscattered current was determined by
taking the difference between the measured current and the current at perfect transmission.
Recently, Comforti et al.\cite{comforti}, have used the three terminal
device consisting of two point contacts shown in Fig. 1.   Consider first the case where the
second point contact (QPC2) is completely open, while the first point contact (QPC1) is weakly
pinched off.  When voltage is applied to lead 1 with leads 2 and 3
grounded, quasiparticles backscattered from QPC1 propagate into lead 3.
This geometry is superior to the two terminal setup for measuring the quasiparticle charge because
the current due to the quasiparticles is isolated in lead 3.  More interestingly, this
may be viewed as a method for generating a {\it dilute beam} of Laughlin quasiparticles  propagating
into lead 3.  This opens the door to experiments that probe the transport properties of and interactions
between individual Laughlin quasiparticles.

\begin{figure}
\epsfxsize=4.in \centerline{\epsffile{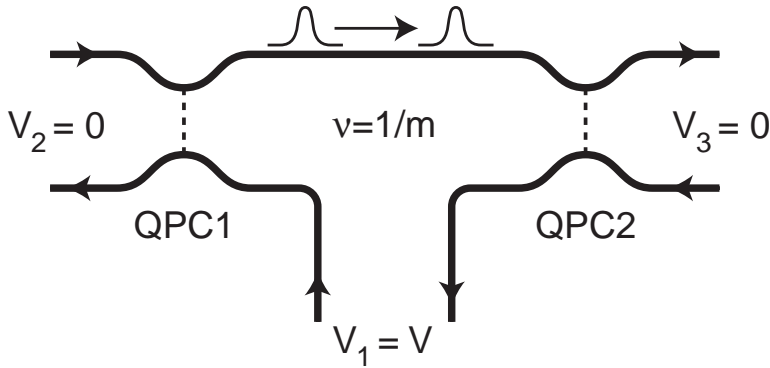}}
\caption{
Schematic diagram of the three terminal fractional quantum Hall device with two quantum
point contacts used by Comforti
et al.  A voltage $V$ is applied to lead 1 with leads 2 and 3 grounded.
When the backscattering at QPC1 is weak, a dilute beam of Laughlin quasiparticles
is directed along the top edge to QPC2.
}
\end{figure}

Comforti et al. \cite{comforti} used this technique to study a dilute beam
of charge $e/3$ quasiparticles after transmission through the second
point contact, QPC2.  By measuring the current and noise in lead 3, they probed the average charge
of the particles transmitted through QPC2.  Surprisingly, they found that even when
the transmission of QPC2 was small, of order $0.1$, the measured transmitted charge $\sim 0.45 e$ was significantly
smaller than that of the electron.  This led them to suggest that perhaps the fractionally charged
quasiparticles in a dilute beam could traverse a nearly opaque barrier.

This suggestion is at odds with the conventional wisdom on the tunneling of quasiparticles.
In the limit of strong pinch off, the quantum Hall fluid is split into two pieces, which each
must have an integer number of electrons.
Coupling them weakly can only give rise
to tunneling of electrons.  Any theory that is perturbative in the tunneling of electrons
will necessarily give noise corresponding to charge $e$.  Nonetheless, it is conceivable that
there could be subtle non perturbative effects.  It is well known that a weakly backscattering
point contact (which is {\it not} two independent quantum Hall fluids)
will cross over at low energy to a regime in which the average current
is well described in terms of the weak tunneling of electrons\cite{kf1,kf2}.
Could the noise in this three terminal setup somehow behave differently?

In this paper we calculate the current and shot noise transmitted
through the QPC2 into lead 3 for the device in Fig. 1.   We employ
the chiral Luttinger liquid model\cite{wen}  with $\nu^{-1}=m$ an odd
integer.  We treat the quasiparticle backscattering from QPC1 at
lowest order in perturbation theory, which guarantees that the
quasiparticles  are dilute and uncorrelated.  For QPC2, we develop
a non perturbative theory, which describes the entire crossover
between the weak and strong backscattering limits.  For the
special case $m=2$ (which does not correspond physically to a FQHE
edge state) we present an exact solution using the technique of
fermionization.  For more general filling factors, $\nu = 1/m$ we
treat the QPC2 perturbatively in the limits of weak tunneling and
weak backscattering. To facilitate comparison with experiments
which are carried out at finite temperature we compute the full
dependence of the current and noise on temperature and voltage.
This gives the crossover between equilibrium noise for $V\ll T$
and shot noise for $V \gg T$.

Our nonperturbative calculation for $\nu = 1/2$
shows that the answer to the question posed above is unambiguously
no.  Fractional charges can {\it not} traverse a nearly opaque barrier.
But the situation is even worse - and more interesting.  We find
that at strictly zero temperature fractional charges cannot even pass through
a nearly perfectly transmitting barrier.  Specifically, the
zero temperature shot noise measured in lead 3 corresponds to charge $e$ particles,
{\it independent} of the transmission of QPC2.
Thus, only electrons are transmitted through QPC2 {\it even when the transmission
of current through QPC2 is nearly perfect}.  We interpret this
result to mean that at zero temperature the transmitted current is
dominated by {\it Andreev} scattering of the incident quasiparticles:
an electron is transmitted, while a hole with the remainder
of the quasiparticle's charge is reflected.

This new and unexpected result points to the subtlety of the zero temperature limit
for fractionalized particles.  When the backscattering at QPC2 is exactly zero,
quasiparticles will obviously be transmitted, and the noise
 in lead 3 should reflect their fractional charge.  Evidently
the limits of taking the temperature to zero and taking the backscattering at QPC2
to zero do not commute.  This situation is unusual in nonequilibrium
many body physics.  Usually, one expects singularities at low energy to be cut off
by both temperature and voltage, with the largest energy scale dominating.  By contrast,
here we have singular behavior in the zero temperature limit for fixed finite
voltage.  While we do not have an exact solution for general filling factors,
our perturbative analysis gives strong evidence that a similar singularity
of the zero temperature limit occurs for $\nu =1/m$.

The outline of the paper is as follows.  In section II we describe
the chiral Luttinger liquid model and establish the notation that we
will use in the remainder of the paper.
The dependence of the current and
noise in lead 3 on temperature, voltage and barrier strength are conveniently described
in terms of scaling functions which are introduced in IIC.

Sections III and IV outline our calculations of the current and
noise.  Readers who are not interested in our methodology can skip
directly to section V where the principle results of those
sections are summarized. In section III we describe our
perturbative analysis.  We begin in IIIA with the simplest limit
in which the backscattering from QPC2 is zero.  In this case the
scaling functions for the current and noise are similar to
previous results for a single junction with a modification due to
the presence of the third lead. In section IIIB we discuss the
large barrier limit, dominated by the tunneling of electrons at
QPC2 and compute the explicit form of the scaling functions for
current and noise as a function of voltage and temperature.  In
IIIC we briefly discuss the perturbation theory for small
backscattering, which has an important divergence in the limit of
zero temperature.
In section IV we describe the exact calculations of the current and noise for $\nu = 1/2$.  We begin
in IVA with a brief discussion of the technique of fermionization and set up the formalism that
we use to calculate the current and noise in IVB and IVC.

Finally, in section V we synthesize the results of sections III
and IV and discuss their implications for experiment.   In section
VA we discuss the scaling behavior of the current and noise as a
function of current and temperature and compare the exact results
for $\nu = 1/2$ with the perturbation theory.    In VB we discuss
the limit of zero temperature and interpret physically the
processes responsible for the singular behavior.  We also propose
a experimental setup to observe this effect.   Finally in VC we
 discuss our results in light of the recent experiments of Comforti et
 al.\cite{comforti}

The calculations presented in this paper were quite involved.  We have relegated
many of the details to two appendices.  In appendix A we discuss our method for evaluating
the correlation functions which arise in our perturbative expansions.
These calculations require a generalization of the Keldysh technique for evaluating non
equilibrium Green's functions.   Many of our results
involve complicated integrals, which are evaluated in appendix B.

\section{Model and Scaling Behavior}

\subsection{Model}

The device in Fig. 1 is described using the chiral Luttinger
liquid model\cite{wen}.  This describes the low energy excitations of the
edge states incident from each of the three leads, as well as the
coupling between them at QPC1 and QPC2.  The Hamiltonian is given
by ${\cal H} = {\cal H}^0_1 + {\cal H}^0_2 + {\cal H}^0_3 + V_1 +
V_2$.  Here ${\cal H}^0_i$ describes a $\nu = 1/m$ chiral
Luttinger liquid edge state which is incident from lead i:
\begin{equation}
{\cal H}^0_i = {m v_F\over {4 \pi }} \int d x_i
[\partial_x\phi_i(x_i)]^2 .
\end{equation}
The coordinates $x_i$
are defined so that at QPC1 $x_i=0$ and at QPC2 $x_i =L$. The
fields $\phi_i(x_i)$ satisfy the commutation relations
$[\phi_i(x_i),\phi_j(x'_j)] =   i (\pi/m)\delta_{ij}  {\rm
sign}(x_i - x'_i)$.   In the following we shall choose units in
which the edge state velocity $v_F = 1$, as well as $\hbar = e =
1$.

Tunneling of charge $1/m$ Laughlin quasiparticles from edge $i=1$
to edge $i=2$ at QPC1 is described by,
\begin{equation}
V_1 = v_1
( O_{1v}^+ e^{-i V t/m} + O_{1v}^- e^{i V t/m}).
\end{equation}
The exponential factors reflect the voltage
difference $V$ between the incident edge states at the junction.
The quasiparticle backscattering operator is given by
\begin{equation} O_{1v}^\pm = {1 \over{(2\pi \eta)^{1/m}}} e^{\pm
i (\phi_1(0) - \phi_2(0))}, \end{equation} where $\eta$ is an
ultraviolet cutoff. QPC2 may similarly be described in terms of
quasiparticle backscattering, \begin{equation} V_{2v} = v_2(
O_{2v}^+  + O_{2v}^-).
 \end{equation}
with \begin{equation}
 O_{2v}^\pm = {1 \over{(2\pi \eta)^{1/m}}}
e^{\pm i (\phi_2(L) - \phi_3(L))}.
\end{equation}

In general, equations (2.3) and
(2.5) should be augmented with Klein factors\cite{klein}, which ensure the correct
commutation relations between $O_{1v}^\pm$ and $O_{2v}^\pm$.  However
in our analysis we will focus on the limit $L\rightarrow \infty$ and
$v_1\rightarrow 0$ (taken {\it before} other limits, such as $T\rightarrow 0$).
In the $L\rightarrow\infty$ limit the Klein factors are unnecessary.

\subsection{Currents and Noise}

Currents can be measured in any of the three contacts.  The
current flowing out contact $i$ is given by the operator,
\begin{equation}
\hat I_i = (\partial_x \phi_{i-1} - \partial_x \phi_i)/2\pi ,
\end{equation}
evaluated at a point in contact i.  (Here $\phi_0$ is identified with
$\phi_3$.)
The measured current will be a function of the voltage $V$ at lead one and
temperature,
and is given by the expectation value,
$I_i(V,T) = \langle \hat I_i \rangle$.  Similarly,
the noise in the limit of zero frequency is\cite{factorof2},
\begin{equation}
S_{ij}(V,T) = {1\over 2} \int dt \left\langle \hat I_i(t) \hat I_j(0) +
\hat I_j(0) \hat I_i(t) \right\rangle.
\end{equation}
For steady state conditions $I_i$ and $S_{ij}$ are independent of the position in the
contact where the current operator is evaluated.

In addition to the noise due to quasiparticles backscattered at
QPC1, $S_{ij}$ will include equilibrium fluctuations in
the current.  The equilibrium fluctuations will be present even when $v_1 = 0$,
though they will
of course be independent of $V$ in that case.  For small $v_1$
the equilibrium noise will be much larger than
the noise due to the backscattered quasiparticles.  We therefore focus on
the {\it excess noise} $\Delta S_{ij}(V,T) = S_{ij}(V,T) - S_{ij}(V=0,T)$.  In our
perturbative expansion of $S$ for small $v_1$, this will be given
by the term second order in $v_1$.

Our main focus in this paper will be on the current and excess noise
transmitted through the second point contact, $I_3(V,T)$ and
$\Delta S_{33}(V,T)$, though in section V we shall briefly discuss the
noise reflected from the second contact $S_{11}(V,T)$ and the
cross correlation $S_{13}(V,T)$.  We will often omit the
subscripts, writing $I_3 = I$ and $\Delta S_{33}= \Delta S$.
The transmitted current and noise give information about the
transparency of QPC2 to the incident beam of dilute quasiparticles
and about the charge of the particles that are transmitted by it.  We define
the effective charge,
\begin{equation}
{\cal Q}(T,V) = \Delta S(V,T)/I(V,T).
\end{equation}
In the limit $V \gg T$, this gives the average charge of the particles
transmitted through the second junction.  If electrons are
transmitted we expect ${\cal Q}(V\gg T) = 1$, while if charge $1/m$
quasiparticles are transmitted we expect ${\cal Q}(V\gg T) = 1/m$.
Moreover, we shall see that for $V \sim T$, ${\cal Q}(V,T)$ has a universal form which
can allow for detailed comparison between experiment and theory.

We also define the {\it transparency}
of QPC2,
\begin{equation}
{\cal T}(V,T) = I(V,T)/I_{\rm in}(V,T)
\end{equation}
where $I_{\rm in}$ is the current incident on QPC2
 along the top edge in Fig. 1, which is equal to
$(e^2/mh)V - I_2$. ($I_{\rm in}(V,T)$ is a property of a single
junction $v_1$.)
${\cal T}$ is small when the second junction is nearly pinched
off while ${\cal T} = 1$ when $v_2=0$ and the transmission is perfect.

\subsection{Scaling Behavior}

A renormalization group analysis shows that the operators
$O_{1,2v}^\pm$ have scaling dimension $1/m$\cite{kf1,kf2}.  It follows that
$v_1$ and $v_2$ both have dimension $1-1/m$.  Provided both $V$ and $T$ are well below the bulk FQHE gap,
the current and noise are expected to satisfy a scaling form:
\begin{equation}
I(V,T) = v_1^2 T^{2/m - 1} \tilde I_m(v_2/T^{1-1/m},V/T)  ,
\end{equation}
\begin{equation}
\Delta S(V,T) = v_1^2 T^{2/m - 1} \tilde S_m(v_2/T^{1-1/m},V/T) ,
\end{equation}
where $\tilde I_m(X,Y)$ and $\tilde S_m(X,Y)$ are {\it universal} functions of both arguments.
Similarly, the effective charge transmitted into lead 3 and the transparency of QPC2 should both scale:
\begin{equation}
{\cal Q}(V,T) = \tilde {\cal Q}_m(v_2/T^{1-1/m},V/T)  ,
\end{equation}
\begin{equation}
{\cal T}(V,T) = \tilde {\cal T}_m(v_2/T^{1-1/m},V/T).
\end{equation}

In the following, we calculate these scaling functions.  In
Section III we
consider the limits $v_2/T^{1-1/m} \rightarrow 0$ and
$v_2/T^{1-1/m} \rightarrow\infty$ where a perturbative analysis is possible.
In section IV we consider the special case $m=2$, where an exact calculation of
these scaling functions is possible.

In addition to computing the shape of the scaling functions, we
find an interesting subtlety in the structure of the scaling
functions when $v_2^{m/(m-1)}, T \ll V$.  To highlight this subtle zero temperature limit
it is useful to consider a slightly different form of
the scaling functions.  Specifically, we define
\begin{equation}
{\cal Q}(V,T) = \tilde {\cal Q}_m'(v_2/V^{1-1/m},V/T) ,
\end{equation}
with similar definitions for $I_i$, $\Delta S_{ij}$ and ${\cal T}$.
The limit $T \rightarrow 0$ is then described by $\tilde
{\cal Q}_m'(v_2/V^{1-1/m},\infty)$.  Interestingly, we find that
this function differs qualitatively from the form of
${\cal Q}_m(v_2/T^{1-1/m},\infty)$.  This difference signifies the fact
that the limits $v_2 \rightarrow 0$ and $T \rightarrow 0$ do
not commute.   We return to this issue in section VA and
discuss in detail its physical meaning.

\section{Perturbation Theory}

In this section we compute the scaling functions $\tilde {\cal Q}_m(v_2/T^{1-1/m},V/T)$
and $\tilde {\cal T}_m(v_2/T^{1-1/m},V/T)$ perturbatively in the limits of large
and small $v_2/T^{1-1/m}$.  We begin with the simplest limit
$v_2=0$, in which the transparency of QPC2 is one.
This will give us the scaling function $\tilde {\cal Q}_m(0,V/T)$.
We then consider the opposite limit $v_2/T^{1-1/m}\gg 1$ which
describes a large barrier and allows us to compute $\tilde {\cal Q}_m(\infty,V/T)$.
 Finally in section IIIC we briefly
discuss the effect of a small, but finite barrier, $0 <
v_2/T^{1-1/m}\ll 1$.

\subsection{Perfect Transmission: $v_2 = 0$}

When $v_2/T^{1-1/m} = 0$, QPC2 becomes
perfectly transmitting.  In this limit, the current and noise should reflect the
quasiparticles backscattered by QPC1.  This is nearly identical to
the single point contact model studied in Refs. \onlinecite{kfnoise,flsnoise},
except for the fact that the current
in lead 3 is only due to the current backscattered at the first
contact.  The remainder of the current exits lead 2.  In appendix A2 we
show how to take this into account.  We find that the current and noise transmitted into lead 3
are given by (2.10, 2.11)
with
\begin{equation}
\tilde I_m(0,V/T) = {1\over {\pi m}}{ |\Gamma\left[1/m + i  V/(2\pi
mT)\right]|^2 \over{ \Gamma(2/m)}} \sinh {V\over{2mT}},
\end{equation}
and
\begin{equation}
\tilde S_m(0,V/T) = {1\over m} \tilde I_m(V/T) \coth{V\over 2mT} - 2 T {\partial \tilde I_m(V/T,0)\over {\partial V}}.
\end{equation}
For $V>>T$ the noise is dominated by the first term in (3.2).  Thus $\tilde {\cal Q}_m(0,V/T\rightarrow \infty) =
1/m$, reflecting the fractional charge of the Laughlin quasiparticles.
For $V \sim T$, thermal fluctuations alter the noise.  Nonetheless, $\tilde {\cal Q}_m(0,V/T)$
has a universal form given by
\begin{equation}
\tilde {\cal Q}_m(0,V/T) = {2\over{\pi m}}{\rm Im}\left[ \psi(1/m + i V/(2 \pi
m T))\right]  ,
\end{equation}
where $\psi(x)$ is the digamma function.  Obviously, $\tilde {\cal
T}_m(0,V/T) = 1$.

\subsection{Large Barriers:  $t_2\rightarrow 0$}

When $v_2/T^{1-1/m} \rightarrow \infty$, QPC2 is
nearly pinched off.  In this limit we expect the noise to reflect
the tunneling of electrons through QPC2.  This may
be described perturbatively using a dual model which describes the
tunneling of electrons with amplitude $t_2$ between two separate quantum Hall fluids\cite{kf1,kf2}.
The Hamiltonian is the same as before with $V_{2v}$
replaced by
\begin{equation}
V_{2t} = t_2 (O_{2t}^+  + O_{2t}^-),
\end{equation}
where the electron tunneling operator is
\begin{equation}
O_{2t}^\pm = {1 \over{(2\pi \alpha)^m}} e^{\pm i m
(\phi_2(L) - \phi_3(L))}.
\end{equation}
The current in the third lead is equal to the tunneling current,
\begin{equation}
\hat I = -i t_2 (O_{2t}^+ - O_{2t}^-).
\end{equation}

The expectation value of the current may be written
\begin{equation}
\left\langle \hat I(t_1) \right\rangle  =  \left\langle T_C \left[ \hat I(\tau_1) e^{-i
\int_C
d\tau (V_{1v}(\tau) + V_{2t}(\tau))} \right] \right\rangle_0   .
\end{equation}
Here $\langle ...\rangle_0$ is a thermal expectation value for $v_1 = t_2 = 0$, and
$V_{1v}$ and $V_{2t}$ are interaction picture operators.
$C$ is the Keldysh contour, which runs from time
$-\infty$ to $\infty$ and then
back to $-\infty$\cite{keldysh}.  $T_C$ specifies time ordering on the Keldysh
contour.  The time $\tau_1 = t_1$ is arbitrary, and can be chosen to lie on the forward
Keldysh path.

We expand to obtain the contribution at order $v_1^2
t_2^2$ and find
\begin{equation} I = {1\over 2}(-i)^3 \int_C d\tau_2 d\tau_3 d\tau_4 \ \left\langle
 T_C \left[\hat I(\tau_1) V_{2t} (\tau_2) V_{1v} (\tau_3) V_{1v}
(\tau_4) \right]\right\rangle_0. \end{equation}
The noise, defined in Eq. (2.7), can similarly be expanded,
\begin{equation}
\Delta S = {1\over 2}(-i)^2 \int dt_2 \int_C d\tau_3 d\tau_4 \ \left\langle T_C \left[\hat
I(\tau_1) \hat I(\tau_2)  V_{1v}(\tau_3) V_{1v}(\tau_4) \right]\right\rangle_0.
\end{equation}
Again, $\tau_{1,2} = t_{1,2}$ can be chosen to lie on the forward
Keldysh path.  We have taken advantage of the symmetry under
interchange of $\tau_1$ and $\tau_2$ to combine the two terms in
(2.7).

Evaluation of the expectation values in (3.8) and (3.9) is complicated
because each time integral has a corresponding sum on the forward
and backward Keldysh paths.  These in turn determine the ordering of the
operators.  In appendix A we describe in detail our method for
handling these sums and evaluating the expectation values.  The
 result is
\begin{equation} I = v_1^2 t_2^2 \int
dt_2 dt_3 dt_4 e^{i V t_{34}/m} \left( G_{2m}^+(t_{12}) -
G_{2m}^-(t_{12})\right) \left[ G_{2/m}^+(t_{34}) (K_1^{-+} - K_1^{--}) +
G_{2/m}^-(t_{34})(K_1^{+-} - K_1^{++})\right]
\end{equation}
and
\begin{equation} \Delta S = v_1^2 t_2^2 \int
dt_2 dt_3 dt_4 e^{i V t_{34}/m} \left( G_{2m}^+(t_{12}) +
G_{2m}^-(t_{12})\right) \left[ G_{2/m}^+(t_{34}) (K_1^{-+} - K_1^{--}) +
G_{2/m}^-(t_{34})(K_1^{+-} - K_1^{++})\right]
\end{equation}
where $t_{ij} = t_i - t_j$,
\begin{equation} G_\alpha^\pm(t) =  \left({T\over{2\sin
\pi T(\eta \pm i t)}}\right)^\alpha  ,
\end{equation}
and
\begin{equation}
K_\alpha^{\sigma_3 \sigma_4} =
\left({\sin \pi T(\eta + i \sigma_3 t_{13})\sin \pi T(\eta + i \sigma_4 t_{24}) \over {
\sin \pi T(\eta + i \sigma_3 t_{23})\sin \pi T(\eta + i \sigma_4
t_{14})}}\right)^\alpha.
\end{equation}

The current and noise are then obtained by substituting (3.12,3.13) into (3.10,3.11).
The results can be cast in the scaling form
\begin{equation}
I(V,T) =  v_1^2 t_2^2 T^{2m + 2/m -3} \tilde I_{t,m}(V/T),
\end{equation}
\begin{equation}
\Delta S(V,T) = v_1^2 t_2^2 T^{2m + 2/m -3} \tilde S_{t,m}(V/T).
\end{equation}
$\tilde I_{t,m}(V/T)$ and $\tilde S_{t,m}(V/T)$ are evaluated in Appendix B.
For the current, the integrals may be evaluated analytically,
giving
 \begin{equation}
\hat I_{t,m}(V/T) =  {1\over \pi} { |\Gamma(m+1/m -1 + i V/(2\pi
m T))|^2 \over{ \Gamma(2m+2/m-2)}}\sinh {V\over{2mT}}.
\end{equation}
$\tilde I_{t,m}(V/T)$ has the limiting behavior,
\begin{equation}
\tilde I_{t,m}(V/T\rightarrow 0)  \propto V/T ,
\end{equation}
\begin{equation}
\tilde I_{t,m}(V/T\rightarrow \infty) =  b_m (V/T)^{2m + 2/m - 3} ,
\end{equation}
with $b_m = (2\pi m)^{3-2/m-2m} /\Gamma(2m+2/m-2)$.

The integrals for the noise are given in Appendix B.1, where they are evaluated analytically
for $m=1$ and $m=2$.  A numerical evaluation of the integrals for $m=3$ is discussed in
section VA.  Here we
focus on the asymptotic behavior,
\begin{equation}
\tilde S_{t,m}(V/T\rightarrow 0) \propto  (V/T)^2 ,
\end{equation}
\begin{equation}
\tilde S_{t,m}(V/T \rightarrow \infty) =  b_m (V/T)^{2m + 2/m -3} ,
\end{equation}
where $b_m$ is the same as in (3.18).

$\tilde I_{t,m}(V/T)$ and $\tilde S_{t,m}(V/T)$ determine the limiting forms of the
scaling functions for transparency and effective charge.
Clearly, $\tilde {\cal T}_m(\infty,V/T) = 0$, and
\begin{equation}
\tilde {\cal Q}_m(\infty,V/T) = \tilde S_{t,m}(V/T)/\tilde I_{t,m}(V/T).
\end{equation}
From (3.18) and (3.20) it is clear that for $V \gg T$
the effective charge is unity,
reflecting the fact that only electrons can traverse a
nearly opaque barrier.

\subsection{Small Barriers:  $v_2\rightarrow 0$}

The presence of small, but finite quasiparticle backscattering
$v_2$ at QPC2 gives rise to a perturbative
correction to the current and noise.  This correction is important
because it contains a divergence which is cut off by the
temperature $T$, but not by the voltage $V$.  This signifies a subtle
non analytic behavior as a function of $v_2$ in the limit of zero temperature.

We consider an expansion of the scaling functions for the
current and noise transmitted into lead 3
in powers of $v_2$:
\begin{equation}
\tilde I_m(v_2/T^{1-1/m},V/T) = \tilde I_m(0,V/T) + {v_2^2\over
T^{2-2/m}} \tilde I_{v,m}(V/T),
\end{equation}
\begin{equation}
\tilde S_m(v_2/T^{1-1/m},V/T) = \tilde S_m(0,V/T) + {v_2^2\over
T^{2-2/m}} \tilde S_{v,m}(V/T).
\end{equation}
The first terms in the expansion were given in Section IIIA.  The
corrections clearly diverge in the limit $V,T\rightarrow 0$ for $m>1$.
This reflects the fact that $v_2$ is a relevant perturbation,
which grows as the energy is lowered.

The scaling functions $\tilde I_{v,m}(V/T)$ and $\tilde S_{v,m}(V/T)$ are
calculated in Appendix A.3 and B.2.  The results are quite unusual.
Usually, one expects a divergence in perturbation theory to
be cut off by the largest available energy in the problem, ${\rm max}(V,T)$.
This would imply that for large $x$, $\tilde S_{v,m}(x) \sim 1/x^{2-2/m}$.
However that is {\it not} the case in the present problem.
We find that $\tilde S_{v,m}(x)$ goes to a {\it constant} at large $x$.
This means that perturbation theory in $v_2$ breaks down for $T\rightarrow 0$
even for fixed finite $V$.

For $v_2^{m/(m-1)} \ll T \ll V$ the effective charge is given by,
\begin{equation} {\cal Q} = {1\over m} + c_m {v_2^2\over
T^{2-2/m}}, \end{equation} where $c_m$ is a positive constant
given explicitly in Appendix B.  Clearly the correction to ${\cal
Q}$ diverges for $T\rightarrow 0$. In the following section we
will show that for $m=2$, ${\cal Q} = 1$ at $T=0$ for arbitrarily
small but finite $v_2$. It is quite likely that this conclusion
holds generally for all values of $m>1$.

\section{Exact Solution for $\nu = 1/2$.}

For intermediate temperatures, $v_2/T^{1-1/m} \sim 1$, calculation of the current and noise
requires a non perturbative technique which is capable of describing the crossover
between the weak and strong barrier limits.  For general $m$ this is
quite difficult, though in principle it should be possible to
adapt the thermodynamic Bethe ansatz which was used by Fendley,
Ludwig and Saleur in their calculation of the current and noise for a single
point contact\cite{flsnoise,flscurrent}.
Here we focus on  the special case $m=2$, where the technique of fermionization
simplifies the problem considerably.  This technique was pioneered
by Guinea\cite{guinea} in the 1980's to solve for the crossover in a model of
dissipative Josephson junctions.  In 1992 we used it
to solve for the crossover between weak and strong barrier
limits for a single impurity in a $g=1/2$ Luttinger liquid, which
determined the universal lineshape of resonances\cite{kf2}.  This
technique was later given a simpler and more elegant reformulation  by
Matveev in a model of strongly coupled quantum dots\cite{matveev}.
A related technique has been applied to the two
channel Kondo problem by Emery and Kivelson\cite{emerykivelson}.

We begin in IVA with a review of the technique of fermionization.  This
will set the stage for the calculation of the current in IVB and the
noise in IVC.

\subsection{Fermionization}

In this section we review the technique of fermionization and set
up the formalism that will be used to calculate the current and noise
in the following sections.
We focus for the moment on the second junction described by the  Hamiltonian
$
{\cal H} = {\cal H}^0_2 + {\cal H}^0_3 + V_2,
$
with $m=2$.
The problem is simplified by transforming to new variables in
which the two channels propagate in the same direction\cite{flscurrent}.
We then transform to sum and difference variables by defining,
\begin{eqnarray}
\phi_\rho(x) &= \phi_2(L+x) + \phi_3(L-x)  , \nonumber\\
\phi_\sigma(x) &= \phi_2(L+x) - \phi_3(L-x).
\end{eqnarray}
These new variables satisfy the commutation relations
$[\phi_a(x),\phi_b(x')] = i\pi \delta_{ab} {\rm sign}(x-x')$ for
$a,b = \sigma,\rho$.
The Hamiltonian is then ${\cal H} =
{\cal H}_\rho + {\cal H}_\sigma$, where
\begin{equation}
{\cal H}_\sigma = \int dx \left\{{1 \over {4\pi}} (\partial_x
\phi_\sigma)^2 +  \delta(x) {v_2 \over\sqrt{2\pi\eta}} 2\cos
\phi_\sigma \right\}.
\end{equation}
${\cal H}_\rho$ is similar, but lacks the second term.
The ``spin" sector ${\cal H}_\sigma$ clearly decouples from the
``charge" sector ${\cal H}_\rho$, and contains all effects
of $V_2$.  The transmitted current operator, $\hat I =
\left[\partial_x\phi_2(x_2 > L) - \partial_x\phi_3(x_3<L)\right]/2\pi$ may
be written in the form
\begin{equation}
\hat I = {1\over 2}(\hat I_{\sigma, \rm in} + \hat I_{\sigma, \rm
out}).
\end{equation}
Here we have defined the incoming and outgoing current operators
$\hat I_{\sigma, \rm in} =  \partial_x\phi_\sigma(x <
0)/2\pi$ and $\hat I_{\sigma, \rm out} =  \partial_x\phi_\sigma(x
> 0)/2\pi$.  In deriving (4.3) we have used the fact that the
corresponding incoming and outgoing currents in the charge sector
are equal in steady state, $I_{\rho, \rm in} = I_{\rho, \rm out}$.

The key observation which makes solution of this problem by
fermionization possible is the fact that the operator
$c(x) = e^{i\phi_\sigma(x)}/\sqrt{2\pi\eta}$ has dimension 1/2 and obeys fermionic
commutation relations $\{c(x),c^\dagger(x')\} = \delta(x-x')$\cite{guinea}.  Directly fermionizing, however,
 leads to a Hamiltonian with a term linear in a fermionic operator, which
 is difficult to analyze.  Following Matveev\cite{matveev} we introduce an auxiliary
 fermionic operator $a$, and defined $\psi(x) =(a + a^\dagger) c(x)$.
 It is straightforward to show that $\{\psi(x),\psi^\dagger(x')\} =
 \delta(x-x')$, so that $\psi(x)$ is also a fermionic operator.
 With this substitution, the fermionized Hamiltonian is quadradic in
 fermion operators,
\begin{equation}
{\cal H}_\sigma =  \int dx \left\{ -i \psi^\dagger \partial_x\psi + v_2
\delta(x)\left[(a+a^\dagger)\psi(x) +
\psi^\dagger(x)(a+a^\dagger)\right]\right\}.
\end{equation}
This Hamiltonian describes a scattering problem in which
fermions incident from $x<0$ scatter from an ``impurity" at $x=0$.
Due to the anomalous terms in the impurity interaction
the fermion can either be transmitted, or Andreev scattered.
${\cal H}_\sigma$ can be diagonalized and written
in a basis of scattering states.
To this end we consider the Heisenberg equations of motion,
\begin{equation}
i \partial_t \psi(x)  = -i\partial_x \psi(x) + v_2 \delta(x) (a + a^\dagger) ,
\end{equation}
\begin{equation}
i \partial_t a   =  v_2 (\psi(0) - \psi^\dagger(0)).
\end{equation}
Scattering state solutions are found by choosing
$\psi(x,t)  = \psi_{k, \rm in} e^{i k( x-t)}/L^{1/2}$  for $x<0$
and $\psi(x,t)  = \psi_{k, \rm out} e^{i k( x-t)}/L^{1/2}$  for $x>0$
with
\begin{equation}
\psi_{k,\rm out} = t_k \psi_{k,\rm in} + r_k \psi^\dagger_{-k,\rm
in}.
\end{equation}
Substituting into (4.5,4.6) and eliminating $a$, the equations are solved when,
\begin{equation}
t_k = {k\over {k + 2 i v_2^2}}; \quad\quad
r_k = {2 i v_2^2 \over {k + 2 i v_2^2}}.
\end{equation}
Here $t_k$ and $r_k$
can be interpreted as the amplitudes for transmission and Andreev
scattering of the incident fermions.

The incident and outgoing currents have the form, $\hat I_{\sigma,
\rm in/out} =  \int (dk/2\pi) : \psi_{k, \rm in/out} ^\dagger \psi_{k, \rm
in/out}:$.
Thus, using (4.3) and (4.7) the current operator may be written
\begin{equation}
\hat I = {1\over 2} \int {dk\over{2\pi}}\left[ |t_k|^2 (\psi^\dagger_{k, \rm in} \psi_{k, \rm in}
-\psi_{-k, \rm in} \psi^\dagger_{-k, \rm in})
+ i |t_k||r_k| (\psi^\dagger_{k, \rm in}\psi^\dagger_{- k, \rm in}
-  \psi_{-k, \rm in}  \psi_{k, \rm in} )\right]  .
\end{equation}
Equation (4.9) expresses the operator for the current transmitted through the second junction in terms
of an operator which acts only on the incident edge states.  The
expectation value of the current can thus be expressed in terms of
a single particle correlation function for the incident particles.  The
noise will be expressed in terms of a two particle correlation function.
In sections IVB and IVC we will calculate the correlation
functions
perturbatively in $v_1$, allowing for a full solution of the
current and noise as a function of $v_2$.

The correlation functions can be evaluated by computing correlations
in the channels incident on the second junction, pretending that
the second junction is not present.  To this end we transform
back to the original bosonic variables $\phi_{2,3}(x)$ by writing
\begin{equation}
(a+a^\dagger) \psi^\dagger_{k, \rm in} = L^{-1/2} \int dx O_{2v}^+(x) e^{-ikx} ,
\end{equation}
with
\begin{equation}
O_{2v}^\pm(x) = {1\over\sqrt{2\pi\eta}} e^{\pm i(\phi_2(L+x)
- \phi_3(L-x))}.
\end{equation}
The current (4.9) may then be rewritten as
\begin{eqnarray}
\hat I = {1\over 2L} \int dx_1 dx_2  & \Bigl[d_1(x_1-x_2)
( O_{2v}^+ (x_1)  O_{2v}^-(x_2) -O_{2v}^-(x_2) O_{2v}^+(x_1)) \nonumber\\
 & +d_2(x_1-x_2) (  O_{2v}^+(x_1) O_{2v}^+(x_2)-O_{2v}^-(x_1) O_{2v}^-(x_2)) \Bigr],
\end{eqnarray}
where
\begin{equation}
d_1(x) = \delta(x)-v_2^2 e^{-2 v_2^2|x|},
\end{equation}
\begin{equation}
d_2(x) = - {\rm
sign}(x) v_2^2 e^{-2 v_2^2|x|} ,
\end{equation}
 are the Fourier transforms of
$|t_k|^2$ and $i |t_k||r_k|$.  Since
$(a+a^\dagger)^2=1$ the auxiliary fermions do not enter into (4.12).
The factor of $L$ in the denominator is present because we have really calculated the
integral of the current over length, $\hat I = L^{-1} \int dx \hat I(x)$.  The $L$ in the denominator will  be
cancelled by an integral over a variable upon
which the integrand does
not depend.

\subsection{Current}

In this subsection we evaluate the current $I = \langle \hat I
\rangle$ perturbatively in $v_1$ using (4.12).  In this case, the
anomalous terms give no contribution.  We thus write
\begin{equation}
I = {1\over {2L}} \int dx_1 dx_2 d_1(x_1-x_2)
A(x_1,x_2)  ,
\end{equation}
with $d_1(x)$ given in (4.13) and
\begin{equation}
A(x_1,x_2)  = \left\langle T_C \left[( O_{2v}^+ (x_1)  O_{2v}^-(x_2) -O_{2v}^-(x_2) O_{2v}^+(x_1))
e^{-i \int_C d\tau V_1(\tau)}\right] \right\rangle_0 .
\end{equation}
Here $\langle ... \rangle_0$ is the thermal expectation value with $v_1=0$.  The time
integral is on the Keldysh contour, and $T_C$ specifies time
ordering on that contour.  Expanding and keeping only the term of
order $v_1^2$ we then find,
\begin{equation}
A(x_1,x_2) = {1\over 2}(-i)^2 \int_C d\tau_3 d\tau_4 \left\langle T_C \left[
( O_{2v}^+ (x_1)  O_{2v}^-(x_2) -O_{2v}^-(x_2) O_{2v}^+(x_1))
V_{1v}(\tau_3)V_{1v}(\tau_4)\right] \right\rangle_0  .
\end{equation}
This has a similar structure to the perturbation theory for the
current outlined in section IIIB.  As in that section we defer to
Appendix A a
discussion of our method for handling the sums over Keldysh paths
and the evaluation of the matrix elements.  The result is
\begin{equation}
A(x_1,x_2) =   v_1^2 \int dt_3 dt_4 e^{iV t_{34}/2}
 \left(G_1^+(x_{12}) - G_1^-(x_{12})\right) \left[G_1^+(t_{34})(K_{1/2}^{-+}-K_{1/2}^{--}) +
G_1^-(t_{34})(K_{1/2}^{+-}-K_{1/2}^{++})\right],
\end{equation}
where $G_1^\pm(t)$ and $K_{1/2}^{\sigma_3\sigma_4}$ are given in section IIIB with $t_{1,2}$ replaced
by $x_{1,2}$.

The current is then obtained by substituting (4.18) into (4.15).
The result can be put into the scaling form
\begin{eqnarray}
I(V,T,v_2) & = v_1^2 \tilde I_2(v_2/T^{1/2},V/T)\\
& = v_1^2 \tilde I_2'(v_2/V^{1/2},V/T).
\end{eqnarray}
The general form of $\tilde I_2(v_2/T^{1/2},V/T)$ may be found in Appendix B.
It a three dimensional integral which can not be evaluated
analytically.  A numerical evaluation of the integral is discussed in Section
V.  Here we focus on limiting behavior, where analytic solution is possible.

In the limit of perfect transmission $v_2/T^{1/2}\rightarrow 0$, we find
\begin{equation}
\tilde I_2(0,V/T) = {1\over 2} \tanh {V\over 4T}.
\end{equation}
This agrees precisely with the result of section IIIA, Eq. (3.1).

In the large barrier limit $v_2/T^{1/2}\rightarrow \infty$ we find
\begin{equation}
\tilde I_2(v_2/T^{1/2}\rightarrow\infty,V/T) = {V^2 + 4\pi^2
T^2\over {128 v_2^4}} \tanh {V\over 4T}  .
\end{equation}
This agrees precisely with the small $t_2$ perturbation theory for
$m=2$ (Eq. 3.16) given the identification $t_2 = \pi/(2 v_2^2)$.

In the limit of zero temperature analytic solution is also possible.  In this case
 it is better to use the scaling function $\tilde I_2'$,
and we find
\begin{equation}
\tilde I_2'(v_2/V^{1/2},\infty) = {1\over 2}\left(1 - {2\over \pi} {\bf
K}(-{V^2 \over{16 v_2^4}})\right)  ,
\end{equation}
where ${\bf K}$ is the elliptic integral of the second kind.  This
function shows a cross over between the large barrier limit,
$\tilde I_2'(v_2/V^{1/2} \rightarrow \infty,\infty) = V^2/(128 v_2^4)$ and the
small barrier limit
$\tilde I_2'(v_2/V^{1/2} \rightarrow 0,\infty) = 1/2 -  [4 v_2^2/(\pi V)] \ln (4V/v_2^2)$.
Note the non analytic behavior of the limit
$v_2\rightarrow 0$.

\subsection{Noise}

The noise is evaluated using equations (2.7) and (4.12).  By shifting
variables $x_k \rightarrow x_k-t$ the integral in (2.7) becomes independent of $t$.
The integral over $t$ then cancels one factor of $L$ and we find
\begin{equation}
\Delta S = {1\over 4L}  \int dx_1 dx_2 dx_3 dx_4 \left[ d_1(x_{12})
d_1(x_{34}) A_1(\{x_k\}) - d_2(x_{13}) d_2(x_{24}) A_2(\{x_k\})
\right],
\end{equation}
where $d_1(x)$ and $d_2(x)$ are given in (4.13,4.14) and
\begin{equation}
A_1(\{x_k\}) =
\Bigl\langle ( O_{2v}^+ (x_1)  O_{2v}^-(x_2) -O_{2v}^-(x_2) O_{2v}^+(x_1))
( O_{2v}^+ (x_3)  O_{2v}^-(x_4) - O_{2v}^-(x_4) O_{2v}^+(x_3))\Bigr\rangle  ,
\end{equation}
\begin{equation}
A_2(\{x_k\}) = \Bigl\langle   O_{2v}^+(x_1) O_{2v}^+(x_3) O_{2v}^-(x_2)O_{2v}^-(x_4)
+  O_{2v}^-(x_4)O_{2v}^-(x_2)O_{2v}^+(x_3) O_{2v}^+(x_1) \Bigr\rangle.
\end{equation}
In the second term of (4.24) and in (4.26) we have permuted the
dummy variables $x_2 \leftrightarrow x_3$ and $x_1 \leftrightarrow x_4$
to make $x_1$, $x_3$ the arguments of $O^+$
and $x_2$, $x_4$ the arguments of $O^-$.
Again the integral depends on only three
of the four $x_k$.  The remaining integral cancels the $L$ in the
denominator.  The expectation values are expanded to order $v_1^2$ and
evaluated in a manner similar to that
in the previous subsection.  Details of this may be found in Appendix A4, where the
analog of equation (4.18) is derived.

The noise is then obtained by substituting (A.38) into (4.24).
The result can be cast in the scaling form
\begin{eqnarray}
\Delta S(V,T,v_2) & = v_1^2 \tilde S_2(v_2/T^{1/2},V/T)\\
& = v_1^2 \tilde S_2'(v_2/V^{1/2},V/T).
\end{eqnarray}
The general form of $\tilde S_2(v_2/T^{1/2},V/T)$ may be found in Appendix B.
It involves a five dimensional integral which can not be evaluated
analytically.  A numerical evaluation of the integral is discussed in Section
V.  As in Section IVB we focus on limiting behavior, where analytic solution is possible.

In the limit of perfect transmission $v_2/T^{1/2}\rightarrow 0$, we find
\begin{equation}
\tilde S_2(0,V/T) = {1\over 4} \tanh^2 {V\over 4T}.
\end{equation}
This agrees precisely with the result of Section IIIA, Eq. (3.2).

In the large barrier limit $v_2/T^{1/2}\rightarrow \infty$ we find
\begin{equation}
\tilde S_2(v_2/T^{1/2}\rightarrow\infty,V/T) = {V^2 + 4\pi^2
T^2\over {128 v_2^4}} \tanh^2 {V\over 4T}  .
\end{equation}
This agrees precisely with the small $t_2$ perturbation theory for
$m=2$ (Eq. B.4), again
with the identification $t_2 = \pi/(2 v_2^2)$.
This confirms that the small $t_2$ perturbation theory is indeed
correct and that in the large barrier limit of low temperature and voltage only electrons can be
transmitted through the second junction.

In the limit of zero temperature the five dimensional integral can still not be fully evaluated.
However, as explained in Appendix B we have established {\it
numerically} that the noise is equal to
\begin{equation}
\tilde S_2'(v_2/V^{1/2},\infty) = {1\over 2}\left(1 - {2\over \pi} {\bf
K}(-{V^2 \over{16 v_2^4}})\right)  ,
\end{equation}
where ${\bf K}$ is the elliptic integral of the second kind.  This
is precisely equal to the current in Eq. 4.23.

This result is quite surprising because it implies that at zero temperature
the effective charge is
\begin{equation}
\tilde {\cal Q}'(v_2/V^{1/2},\infty) = 1  ,
\end{equation}
{\it independent} of the barrier strength $v_2$.  Thus the shot noise
measured in the third contact indicates that electrons are
transmitted even when $v_2/V^{1/2}$ is small and the transmission
through the second contact is nearly perfect.

\section{Discussion}

We now synthesize the results of the preceding two sections and discuss
their physical meaning and their implications for experiment.  We begin
with a summary of our results for the dependence of the current and noise
on temperature and voltage.  We then discuss in detail the zero temperature
limit, and identify
the processes responsible for the peculiar behavior that occurs
there.  We propose a new experiment to probe the new physics that occurs near zero temperature.
Finally, we discuss the implications of our results for existing
experiments.

\subsection{Current and Noise}

Fig. 2 shows the transparency of QPC2 $\tilde {\cal T}_2(v_2/T^{1/2},V/T)$ and the effective charge
$\tilde {\cal Q}_2(v_2/T^{1/2},V/T)$  for $\nu=1/2$ as a function of $V/T$ for various temperatures.
The lowest temperatures have the smallest transparency and the largest effective charge.
These curves were obtained by evaluating the integrals in appendices B.3 and B.4 numerically.
The thick curves in Fig 2b are the asymptotic results from perturbation theory in the limits
$v_2/T^{1/2}\rightarrow 0$ (Eq. 3.3) and $v_2/T^{1/2}\rightarrow \infty$ (Eq. B6).  In the limit
of low temperature (or large backscattering at QPC2) the results of the exact calculation
reduce to the results of perturbation theory based on the weak tunneling of electrons.  Moreover,
comparing Fig. 2a and 2b, it is clear that when the transparency is small, the effective charge
(for $V/T$ sufficiently large) is very close to $1$.
\begin{figure}
\epsfxsize=7.in
\centerline{\epsffile{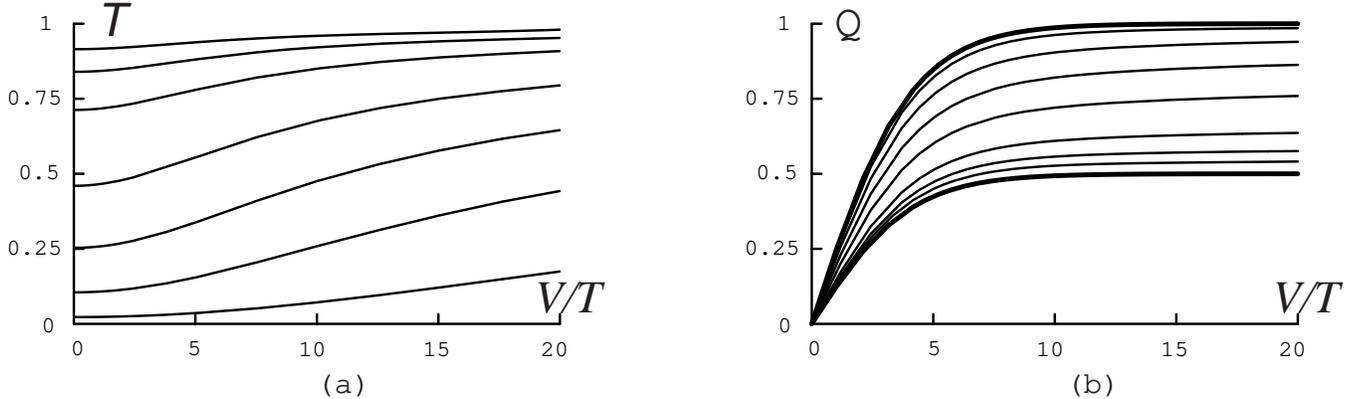}}
\caption{
(a) Transparency of
QPC2, $\tilde {\cal T}_2(v_2/T^{1/2},V/T)$  as a function of $V/T$
for different temperatures, $T/v_2^2 = 0.2,0.5,1,2,5,10,20$.
The lowest temperatures have the lowest transparency.
(b) Effective charge transmitted
through QPC2, $\tilde {\cal Q}_2(v_2/T^{1/2},V/T)$ as a function of $V/T$ for the same
set of temperatures as in (a).  The lowest temperatures
have the largest effective charge.  The thick lines are the asymptotic limits
${\cal Q}(0,V/T)$ and ${\cal Q}(\infty,V/T)$ discussed in the text. }
\end{figure}

A striking feature of these curves is their behavior for large $V/T$.  For each of the curves
in Fig. 2a the transparency increases with increasing $V/T$ and eventually approaches 1.  This
is because the transmission through QPC2 becomes perfect for $V \gg v_2^2$.  By contrast, the
curves for the effective charge in Fig. 2b saturate at a constant value
 for $V/T\rightarrow \infty$.
Thus, even though the transmission through QPC2
$\tilde {\cal T}_2(v_2/T^{1/2}, \infty)=1$ is perfect, the charge
$\tilde {\cal Q}_2(v_2/T^{1/2}, \infty)$ of the transmitted particles
is {\it not} equal to the charge $1/2$ of the
quasiparticles incident on QPC2, but rather varies between $1/2$ and $1$
as the temperature is lowered.  In striking contrast, at zero temperature
equations (4.23) and (4.31) show that the effective charge of the transmitted particles
$\tilde {\cal Q}_2'(v_2/V^{1/2}, \infty)=1$,  independent of the voltage $V$.  The scaling
functions $\tilde {\cal Q}_2(v_2/T^{1/2}, V/T)$ and $\tilde {\cal Q}_2'(v_2/V^{1/2}, V/T)$ thus show
qualitatively different behavior.  This is quite unusual, since usually the dependence of
scaling functions on voltage and temperature are qualitatively similar.
The origin of this behavior can be traced to the singular behavior of limit $T\rightarrow 0$
with fixed $V$:
\begin{eqnarray}
{\cal Q}(v_2=0,T\rightarrow 0,V) &= 1/2 , \nonumber\\
{\cal Q}(v_2\rightarrow 0,T = 0,V) &= 1.
\end{eqnarray}
The limits of $T\rightarrow 0$ and $v_2\rightarrow 0$ do not commute.

In section VB we will offer a physical interpretation of this peculiar behavior.  However, before doing
so it is important to ask whether it is an artifact of the chiral edge theory for
$\nu= 1/2$, or whether it also occurs more generally.  In Fig. 3 we show perturbative calculations
of $\tilde {\cal Q}_3(0,V/T)$ (Eq. 3.3) and $\tilde {\cal Q}_3(\infty,V/T)$ (Eqs. 3.21, 3.16, B.4).  It seems
quite plausible that for intermediate temperatures $\tilde {\cal Q}_3(v_2/T^{2/3},V/T)$ should interpolate
smoothly between the two limits as in Fig. 2b.  This does not exclude the possibility, however,
that the curves cross over to $1/3$ for $V/T \gg v_2^{3/2}/T$.  This is ruled out, however, by
perturbation theory in $v_2$.   Eq. 3.24 shows that for $v_2^{3/2} \ll T \ll V$,
$\tilde {\cal Q} - 1/3 \propto v_2^2/T^{4/3}$.  Thus, $\tilde {\cal Q}_3(v_2/T^{2/3},\infty)>1/3$
for finite $v_2$, and it presumably then
crosses over to $1$ for $v_2 \gg T^{2/3}$.

\begin{figure}
\epsfxsize=3.in
\centerline{\epsffile{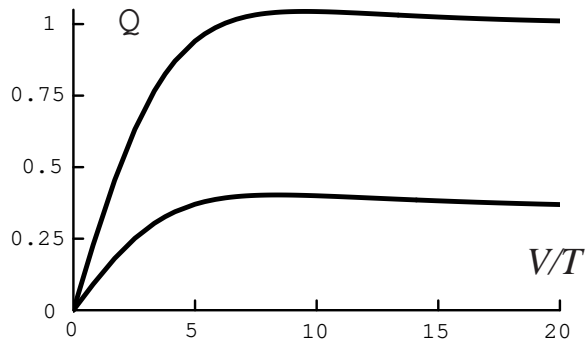}}
\caption{
Scaling functions for the effective charge transmitted into lead 3
for $\nu=1/3$, $\tilde{\cal Q}_3(v_2/T^{2/3},V/T)$.  The bottom curve is
in the weak backscattering or high temperature limit, $\tilde{\cal Q}_3(0,V/T)$,
whereas the top curve is in the low temperature limit, $\tilde{\cal Q}_3(\infty,V/T)$.
Notice that in this limit of an opaque barrier ($v_2 \gg T^{2/3}$) only electrons are transmitted
through QPC2 when $V >> T$.
}
\end{figure}

The small $v_2$ perturbation theory also gives a diverging correction to the transparency
at zero temperature (Eq. 3.22).  This divergence was absent for $\nu = 1/2$.
One may therefore worry that the transparency also goes to zero at $T=0$ for fixed
$V$.  However, this is contradicted by the small $t_2$ perturbation theory (3.14,3.18), which gives
a finite transparency ${\cal T}_3 \propto t_2^2 V^{11/3}$ at $T=0$.  It is most likely that
the divergence for small $v_2$ signifies that transparency is not analytic at $v_2=0$.
Such a non analyticity also occurs for $m=2$, where (4.23) gives
${\cal T}_2 \sim 1 - (v_2^2/V) \log V/v_2^2$.

The above arguments give strong evidence that the singular behavior at $T=0$ that we
have established for $\nu=1/2$ also occurs for $\nu = 1/3$ and other Laughlin filling fractions.
Nonetheless, it would be desirable to obtain a full solution for $\nu =1/m$.
Using the thermodynamic Bethe ansatz, Fendley, Ludwig and Saleur\cite{flsnoise,flscurrent}
have calculated the current
and noise for a single point contact, accounting for the full crossover between the
weak backscattering and strong backcattering limits.  It should be possible
to generalize their formalism to the present 3-terminal geometry.

\subsection{Physical Picture for the $T\rightarrow 0$ limit }

In this section we attempt to make sense out of the peculiar behavior we have
established at zero temperature.  We wish to understand how electrons can
be transmitted through QPC2 into lead 3 even when the transparency of QPC2 is nearly
perfect.  We assume here that this effect occurs for $\nu=1/m$.

At zero temperature quasiparticles backscattered by QPC1 come in
wave packets of charge $e/m$ and duration $\sim 1/V$ at a rate
$\sim v_1^2 V^{(2/m)-1}$.  For $v_1 \ll V^{1-(1/m)}$ the quasiparticle wave packets
are independent and can be considered one at a time.
The interaction of a quasiparticle with QPC2 presents a scattering problem.
When a quasiparticle scatters from QPC2 it is natural to ask what comes out.
Unlike the non interacting electron version of this problem the number of
quasiparticles is not necessarily conserved in this scattering process.
However, the total charge is conserved.
We consider three processes. (1)  The quasiparticle is transmitted with
probability {\sf T} into lead 3.  (2) The quasiparticle is reflected with
probability {\sf R} into lead 1.  (3) The quasiparticle is {\it Andreev reflected}
with probability {\sf A}.  In this process an electron, with charge $e$ is
transmitted into lead 3 while a hole with charge $(-1+1/m)e$ is reflected into lead
1.  It is straightforward to show that if these are the only allowed processes
(i.e. ${\sf T} + {\sf R} + {\sf A} = 1$) the transparency is given by,
\begin{equation}
{\cal T} = {\sf T} + m {\sf A}.
\end{equation}
Moreover, the effective charge will be,
\begin{equation}
{\cal Q} = {{\sf T} + m^2{\sf A} \over {m{\sf T} + m^2{\sf A}}} .
\end{equation}

For $v_2 \ll T^{1-1/m}$ we clearly have ${\sf R}={\sf A} = 0$ and ${\sf T}=1$.
On the other hand, at zero temperature our noise calculation shows that
${\sf T} = 0$, since only electrons were found to be transmitted into lead 3.
The transmitted current is thus apparently
dominated by Andreev processes.
This is no surprise in the large barrier limit $v_2 \gg V^{1-1/m}$, where ${\cal T}$
is small, so that ${\sf A}$ is small and ${\sf R} \sim 1$.  In the small barrier limit
$v_2 \ll V^{1-1/m}$, however, we have ${\cal T} \sim 1$.  This then implies that
${\sf A} = 1/m$ and ${\sf R} = 1-1/m$.  Thus, quite remarkably, the incident quasiparticle is either reflected
or Andreev reflected with probabilities that have saturated at values which conspire to
give {\it perfect}
transmission of the {\it current}.  Moreover, in this limit the time averaged
current backscattered off QPC2 {\it vanishes}, although it will be noisy as we now detail.

A key feature of the Andreev processes is that the transmitted and reflected currents are
{\it correlated}.  These correlations give an unambiguous signature
in the noise.  We therefore propose that the noise be measured in {\it both} leads 1 and 3.
It may be desirable
to add an additional lead between leads 1 and 3
which can isolate the current reflected at QPC2.  In any case, this will
not affect the following zero temperature predictions.
As above, the noise measured in lead 3 should
reflect the charge $e$ of the Andreev transmitted electrons,
\begin{equation}
\Delta S_{33}  = I_3.
\end{equation}
The noise measured in lead 1, however, will be a combination
of the charge $1/m$ reflected quasiparticles and the charge $(1/m)-1$ Andreev reflected holes.
In terms of the measured currents, it will be given by
\begin{equation}
\Delta S_{11}  = (1/m)\Delta I_1 + (1-1/m) I_3 .
\end{equation}
Here $\Delta I_1$ is the current flowing into lead 1 due to the reflections from QPC2, that is
$\Delta I_1 = I_1+ V e^2/(m h)$.
If an additional lead, say lead 4, is present between leads 1 and 3, then
for $\Delta S_{44}$ one would have simply $\Delta I_1$ replaced by $I_4$ in equation (5.5).
The cross correlations are determined solely by the Andreev processes,
\begin{equation}
\Delta S_{13} = -(1-1/m) I_3.
\end{equation}
In the limit of weak pinch off for QPC2, we have
$\Delta I_1 = 0$, at zero temperature.  Nevertheless, the current flowing into lead 1 is noisy, with
$\Delta S_{11} = -\Delta S_{13} = (1-1/m) \Delta
S_{33}$.  Thus, in this way one can prepare a noisy but zero time-averaged
non-equilibrium current, present in the zero temperature limit
where equilibrium current fluctuations vanish.  While undoubtedly challenging,
it would be fascinating to detect this effect and the presence of
Andreev processes more generally.

\subsection{Relation to Existing Experiments}

We close by commenting briefly on the implications of our results for the
experiments of Comforti et al.\cite{comforti}.  It is clear that our results give no support
to the notion of fractional charges traversing a nearly opaque barrier.
So the interpretation of the data remains a puzzle.  However, it is worthwhile to
point out some possible sources of discrepancy.

The exact scaling functions for
$m=2$ which we have computed are strictly speaking only applicable
for a point contact which backscatters
high energy
(but still below the bulk FQHE gap)
incident particles only weakly.
A point contact which is strongly pinched off will not generally follow
the universal
crossover between weak and strong backscattering embodied in the scaling functions.
Nevertheless, since our results
show an absence of any
subtle non-perturbative effects in the limit of weak tunneling through the point contact,
it is difficult to imagine
that this could modify our basic conclusion that only electrons can traverse an opaque barrier.
It seems plausible
that the experimentally
observed charge of $0.45$ is a finite
temperature crossover effect,
which might well
revert to a charge of $e$ as the temperature is lowered further.
But it remains difficult reconciling a
transmitted
charge well below the electron charge
for a point contact with such a small measured
transparency of only
$0.1$.

\begin{figure}
\epsfxsize=3.in
\centerline{\epsffile{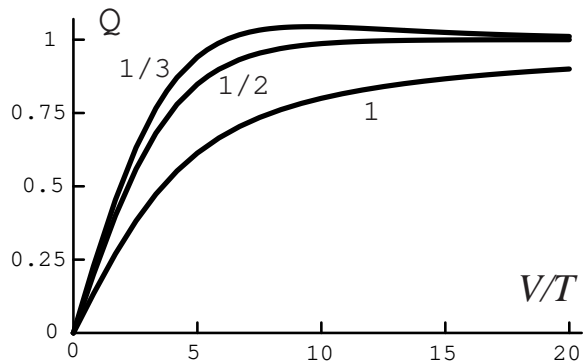}}
\caption{
Scaling functions $\tilde{\cal Q}_m(v_2/T^{1-1/m},V/T)$
for the effective charge transmitted into lead 3
through QPC2 in the large barrier limit, $v_2/T^{1-1/m} \rightarrow \infty$.
The three curves correspond to filling $\nu = 1/m = 1, 1/2, 1/3$ as labelled.
}
\end{figure}

Comforti et al. \cite{comforti} extracted the effective charge by fitting the measured $I(V,T)$ and
$\Delta S(V,T)$ to an ``independent particle model", which is essentially the non interacting electron
version ($m=1$) of the scaling functions $\tilde I_m(V/T)$ and $\tilde S_m$\cite{lesovik}.
In Fig. 4 we compare the scaling functions for the effective charge in the large
barrier limit $\tilde {\cal Q}_m(\infty,T/V)$ for $m=1$, $2$ and $3$.  Here
$\tilde {\cal Q}_1(\infty,x) = \coth x/2 - 2/x$, $\tilde {\cal Q}_2(\infty,x) = \tanh x/4$,
and $\tilde {\cal Q}_3(\infty,x)$ is computed numerically as in Fig. 3.
The curves clearly differ quantitatively.  The results of this paper thus suggest an alternative
method for analyzing the data:  For fixed temperature plot the measured values of
$\Delta S(V,T)/I(V,T)$ as a function of $V/T$ and compare with the scaling functions
$\tilde {\cal Q}_3(0,V/T)$ and $\tilde {\cal Q}_3(\infty,V/T)$ in Fig. 3.  For data taken
at voltages $V \gtrsim 10 T$ conclusions about the asymptotic charge for $V\gg T$ should not
depend on the fitting method.  But for smaller voltages there may well be a difference.

\acknowledgements

It is a pleasure to thank Moty Heiblum for challenging us to think
about this problem and sharing his data prior to publication.
We also thank Andrei Shytov for many helpful
discussions.  C.L.K. wishes to thank the Institute for Theoretical
Physics, where this work was initiated.
M.P.A.F. was generously supported by the NSF under grants
DMR-0210790 and PHY-9907949.

\begin{appendix}

\section{Expectation values and Keldysh sums}

In this appendix we demonstrate our technique for evaluating the
expectation values and sums over Keldysh paths.  In section A1 we
do in detail the calculation for the small $t_2$ limit.  This will establish
our method, which can then be applied to the other calculations.
In A2 we discuss the limit of small $v_2$.  Finally in A3 we briefly discuss the
calculations for the exact current and noise for $m=2$.

\subsection{Small $t_2$ Perturbation Theory}

In this section we provide some details of the calculation which
leads from equation (3.8,3.9) to (3.10,3.11).  Our starting point is the
expansion of  the current and noise to order $v_1^2 t_2^2 $.
It is useful to introduce an index $\sigma = \pm$ which specifies the
forward and backward paths of the Keldysh contour.
Then, $\int_C d\tau \rightarrow \sum_\sigma \sigma \int dt
$.  For the variable $t_1$  (and $t_2$ for the noise)
we introduce a dummy sum over $\sigma_1$ (and $\sigma_2$).
In addition, we write the two terms in the tunneling Hamiltonian (3.4)
and the current operator (3.6) as
a sum over $s = \pm$.  The current and noise can then be written,
\begin{equation}
I = {1\over 4}v_1^2 t_2^2 \sum_{\{\sigma_k,s_k\}}
 s_1 \sigma_2 \sigma_3 \sigma_4 \int d^3 t
 e^{-i V(s_3 t_3 + s_4 t_4)/m } \nonumber
 \left\langle T_C \left[ O^{s_1}_{2t}(\sigma_1 t_1)
 O^{s_2}_{2t}(\sigma_2 t_2) O^{s_3}_{1v}(\sigma_3 t_3)  O^{s_4}_{1v}(\sigma_4 t_4) \right]
 \right\rangle_0
\end{equation}
\begin{equation}
\Delta S = {1\over 8} v_1^2 t_2^2\sum_{\{\sigma_k,s_k\}}
 s_1 s_2 \sigma_3 \sigma_4 \int d^3 t
 e^{-i V(s_3 t_3 + s_4 t_4)/m } \nonumber
\left\langle T_C \left[ O^{s_1}_{2t}(\sigma_1 t_1)
 O^{s_2}_{2t}(\sigma_2 t_2) O^{s_3}_{1v}(\sigma_3 t_3)  O^{s_4}_{1v}(\sigma_4 t_4) \right] \right\rangle_0
\end{equation}
The three time integrals are over $t_2$, $t_3$ and $t_4$.  Note, however,
 that due to invariance with respect to time translations they can be
 shifted to any three of the times $t_1, t_2, t_3, t_4$.
 Clearly we must have $s_1+s_2 = s_3+s_4=0$ in each of the sums on $\{s_k\}$.  By appropriately
 re-labeling the integration variables, we may specify $s_1 = -s_2 = -s_3 =
 s_4 = +$.
 \begin{equation}
I = {1\over 2} v_1^2 t_2^2 \sum_{\{\sigma_k\}}
 (\sigma_2-\sigma_1) \sigma_3 \sigma_4 \int d^3 t
 \Pi(\{\sigma_k , t_k\}) e^{i V t_{34}/m }
\end{equation}
\begin{equation}
\Delta S = {1\over 2} v_1^2 t_2^2 \sum_{\{\sigma_k\}}
\sigma_3 \sigma_4 \int d^3 t
 \Pi(\{\sigma_k , t_k\}) e^{i V t_{34}/m }
\end{equation}
where
\begin{equation}
 \Pi(\{\sigma_k , t_k\}) = \left\langle T_C\left[ O^+_{2t}(\sigma_1 t_1)
 O^-_{2t}(\sigma_2 t_2) O^-_{1v}(\sigma_3 t_3)  O^+_{1v}(\sigma_4 t_4) \right]\right\rangle_0.
 \end{equation}

 $ \Pi(\{\sigma_k , t_k\})$ is computed
 by first computing the imaginary time ordered correlation
function.
\begin{equation}
\Pi(\{\tau_k\}) = \left\langle  T_\tau\left[ O^+_{2t}(\tau_1)
 O^-_{2t}(\tau_2) O^-_{1v}(\tau_3)  O^+_{1v}(\tau_4)\right]\right\rangle_0.
 \end{equation}
The expectation value  factorizes
into three terms,
\begin{equation}
\Pi(\{\tau_k\}) =
{\left\langle T_\tau\left[ e^{i(\phi_1(0,\tau_3) -\phi_1(0,\tau_4))}\right]\right\rangle_0
\left\langle T_\tau\left[ e^{i(\phi_3(L,\tau_1) -\phi_3(L,\tau_2))}\right]\right\rangle_0
\left\langle T_\tau\left[ e^{i(\phi_2(L,\tau_1) -\phi_2(L,\tau_2)-\phi_2(0,\tau_3) +
\phi_2(0,\tau_4))}\right]\right\rangle_0\over
{(2\pi\eta)^{2m+2/m}}}
\end{equation}
where $T_\tau$ signifies time ordering in imaginary time.
Using the Hamiltonian (2.1) it is straightforward to show that
\begin{equation} \Pi(\{\tau_k\}) = {  (T/2)^{2m+2/m}
\over {
\sin^{2m}\pi T (\eta + \sigma_{12}\tau_{12})
\sin^{2/m}\pi T (\eta + \sigma_{34} \tau_{34})}}
{ \sin\pi T (\eta + \sigma_{13}(\tau_{13}-i L))
\sin\pi T (\eta + \sigma_{24}(\tau_{24}- i L))\over {\sin\pi T (\eta + \sigma_{23}(\tau_{23}-i L))\sin\pi T
(\eta + \sigma_{14}( \tau_{14}- i L))}}.
\end{equation}
Here $\sigma_{ij} = {\rm sign}(\tau_i-\tau_j)$ reflects ordering of the operators in the imaginary
time ordered product.

The real time correlation functions are determined by taking
$\tau_{ij} \rightarrow  i  t_{ij}$.   The operator ordering is now determined
by the time ordering on the Keldysh contour.  Thus
$\sigma_{ij} = \pm 1$ depending on whether the time $t_i \sigma_i$
comes later or earlier than $t_j \sigma_j$ on the Keldysh contour.
$\sigma_{ij}$ now depends on the Keldysh paths $\sigma_i$, $\sigma_j$
as well as the sign of the time difference $s_{ij} = {\rm sign}(t_i-t_j)$.
Explicitly it may be written
\begin{equation}
\sigma_{ij} = {1\over 2}\left[(\sigma_j - \sigma_i) + s_{ij}
(\sigma_i + \sigma_j)\right].
\end{equation}

In the limit of large $L$ the only times to
contribute will be those with $t_{1,2} \sim t_{3,4}+L$.
Therefore from (A10), $\sigma_{13} = \sigma_{23} = \sigma_3$ and $\sigma_{14}
= \sigma_{24} = \sigma_4$.
The real time correlation function may then be expressed in the form
 \begin{equation}
 \Pi(\{\sigma_k,t_k\}) = G^{\sigma_{12}}_{2m}(t_{12})
 G^{\sigma_{34}}_{2/m}(t_{34})K_1^{\sigma_3 \sigma_4}
 \end{equation}
 with
\begin{equation} G_\alpha^\pm(t) =  \left({T\over{2\sin
\pi T(\eta \pm i t)}}\right)^\alpha
\end{equation}
and
\begin{equation}
K^{\sigma_3 \sigma_4}_\alpha =
\left({\sin \pi T(\eta + i \sigma_3 (t_{13}-L))\sin \pi T(\eta + i \sigma_4 (t_{24}-L)) \over {
\sin \pi T(\eta + i \sigma_3( t_{23}-L))\sin \pi T(\eta + i \sigma_4
(t_{14}-L))}}\right)^\alpha.
\end{equation}
$ G_\alpha^\pm(t)$ may be interpreted as a two point Green's
function more commonly referred to as $G^{<,>}(t)$.  For instance
$ G_{2m}^+(t) = \left\langle  O_{2t}^+(t) O_{2t}^-(0) \right\rangle_0$ and
$ G_{2m}^-(t) = \left\langle  O_{2t}^-(0) O_{2t}^+(t) \right\rangle_0$.

Substituting (A10-A12) into (A3) the sums on $\{\sigma_k\}$ may be
evaluated giving
\begin{eqnarray}
I =  v_1^2 t_2^2 \int d^3 t e^{iV t_{34}/m} (G^+_{2m}(t_{12})-G^-_{2m}(t_{12}))\biggl[
&G^+_{2/m}(t_{34})\left( K_1^{-+} + {1\over 2}(K_1^{++} + K_1^{--})
+ {1\over 2} s_{34}(K_1^{++} - K_1^{--})\right) \nonumber\\
+ &G^-_{2/m}(t_{34})\left( K_1^{+-} + {1\over 2}(K_1^{++} + K_1^{--})
- {1\over 2} s_{34}(K_1^{++} - K_1^{--})\right) \biggr].
\end{eqnarray}
This equation may be simplified by considering the dependence of the integrand on
the ``average time difference" $t_0 = (t_1+t_2 - t_3 - t_4)/2$.  $t_0$ enters the
only in the form $L \rightarrow L - t_0$ and may be
interpreted as the time it takes quasiparticles to propagate
between the two junctions.  It can be shown by contour integration that
\begin{equation}
\int dt_0 (K^{++} - K^{--} ) = 0.
\end{equation}
This allows us to rewrite (A13) in the simpler form,
\begin{equation} I =  v_1^2 t_2^2 \int
dt_2 dt_3 dt_4 e^{i V t_{34}/m} \left( G_{2m}^+(t_{12}) -
G_{2m}^-(t_{12})\right) \left[ G_{2/m}^+(t_{34}) (K_1^{-+} - K_1^{--}) +
G_{2/m}^-(t_{34})(K_1^{+-} - K_1^{++})\right].
\end{equation}
The sum over $\sigma_k$ for the noise is almost the same, except
for the first term in (A3).  This gives
\begin{equation} \Delta S =  v_1^2 t_2^2 \int
dt_2 dt_3 dt_4 e^{i V t_{34}/m} \left( G_{2m}^+(t_{12}) +
G_{2m}^-(t_{12})\right) \left[ G_{2/m}^+(t_{34}) (K_1^{-+} - K_1^{--}) +
G_{2/m}^-(t_{34})(K_1^{+-} - K_1^{++})\right].
\end{equation}
Finally, in equations (3.10,3.11) we have shifted $t_{1,2}
\rightarrow t_{1,2}-L$ to eliminate the variable $L$.

\subsection{Current and Noise for $v_2  = 0$}

Here we briefly outline the calculation of the current and noise
when the second junction transmits perfectly.  Similar results have
been obtained in earlier for the single junction\cite{kfnoise,kf1,kf2}.  We include the
calculation here because the result is slightly different and
because we use a somewhat different method, which will be useful when
generalizing to finite $v_2$.

We express
the current $I=I_3$ in terms of incident currents and the current
backscattered at the first junction,
\begin{equation}
\hat I = \hat I_{2 \rm in} - \hat I_{3 \rm in} + \hat I_{b1}
\end{equation}
where $I_{2,3 \rm in} = \partial_x \phi_{2,3}/2\pi$ are the currents carried by the chiral edge
states incident from leads 2 and 3.  The current backsattered at
the first junction is
\begin{equation}
\hat I_{b1} = -i( v_1/m)( O_{1v}^+ e^{-i V t/m} - O_{1v}^- e^{i V t/m}).
\end{equation}
The backscattered current is related to the voltage drop across
the junction, discussed in Ref. \onlinecite{kfnoise}.
For the current the first two terms in (A18) cancel, and we have $I =
\langle I_{b1} \rangle$.  This may be evaluated using the procedure in appendix
A1 to be
\begin{equation}
I = {v_1^2\over m} \int dt e^{iVt/m} \left( G^+_{2/m}(t) - G^-_{2/m}(t)
\right)
\end{equation}
with $G^\pm_{2/m}(t)$ given in (A11).  Evaluation of the integral gives the result quoted in (3.1).

The excess noise contains two contributions,
\begin{equation}
\Delta S = \Delta S_{b1,b1} + 2 \Delta S_{b1,2 \rm in}.
\end{equation}
The fluctuation in the backscattered current
$\Delta S_{b1,b1} = (1/2)\int dt \langle \hat I_{b1}(t) \hat I_{b1}(0)
\rangle$ is related to the voltage fluctuations across the junction.
It has the form (check sign),
\begin{equation}
\Delta S_{b1,b1} ={ v_1^2\over m^2} \int dt e^{iVt/m} \left( G^+_{2/m}(t) +
G^-_{2/m}(t)\right).
\end{equation}
Using the fact that $G^-(t + i/T) = G^+(t)$ it is straightforward
to establish that $\delta S_{bb} = (I/m) \coth V/2mT$.  Physically,
the two terms in (A19) and (A21) describe the rates for forward and backward
tunneling of quasiparticles across the voltage difference $V$
which are related by a factor $e^{V/mT}$.

The second term,
$\Delta S_{b1,2\rm in} = (1/2)\int dt \langle \{\hat I_{b1}(t), \hat I_{2 \rm in}(0) \}
\rangle$ gives the cross correlation between the backscattered
current and thermal fluctuations in the current incident from lead 2.
This cross correlation can be shown to have the form
\begin{equation}
\Delta S_{b1,2 \rm in} =  T {\partial \langle I_{b1} \rangle \over{\partial
V_2}}
\end{equation}
In equilibrium, $V \rightarrow 0$ this is simply a statement of the fluctuation
dissipation theorem.  However, as shown in Ref. \onlinecite{kfnoise}, it is also valid for $V>0$.

Combining (A21) and (A22) we get the result quoted in (3.2). Note that
the other terms present in $S$ do not contribute to the excess noise.  In
particular the current incident from lead 3 will have no correlation with $I_b$.

\subsection{Small $v_2$ perturbation theory}

When $v_2$ is finite we write the current as
\begin{equation}
\hat I = I_{2, \rm in} - I_{3, \rm in}+I_{b1}- I_{b2},
\end{equation}
where the current backscattered at the second junction is
\begin{equation}
\hat I_{b2} = -i( v_2/m)( O_{2v}^+ - O_{2v}^- ).
\end{equation}

The average current at order $v_2^2$ is then given by $I = I_{b2} $.
This may be computed along the lines of the previous section.  The structure is
almost identical to (A15), except the dimensions of the operators is changed.  We
find
\begin{equation} I_{b2} = { v_1^2 v_2^2\over m} \int
d^3 t e^{i( V_{12} t_{34} - V_{32} t_{12})/m} \left( G_{2/m}^+(t_{12}) -
G_{2/m}^-(t_{12})\right) \left[ G_{2/m}^+(t_{34}) (K_{1/m}^{-+} - K_{1/m}^{--}) +
G_{2/m}^-(t_{34})(K_{1/m}^{+-} - K_{1/m}^{++})\right].
\end{equation}
For  use in the next section we have included voltages $V_k$ in all three contacts,
and $V_{kl} = V_k-V_l$.  The current is evaluated with $V_2=V_3=0$ and $V_1=V$.

From (A23), the nonzero  contributions to the excess noise at order $v_2^2$ will be given by
\begin{equation}
\Delta S = \Delta S_{b2,b2} -2 \Delta S_{b1,b2} - 2\Delta S_{b2,2 \rm in} + 2 \Delta S_{b2,3 \rm in}
\end{equation}
As in the previous section, the cross correlations with the incident currents have the form
\begin{equation}
\Delta S_{b2,k \rm in} =  T {\partial I_{b2}\over {\partial V_k}}
\end{equation}
for $k=1,2$.  In addition we find
\begin{equation} \Delta S_{b2,b2} = {v_1^2 v_2^2 \over m^2} \int
d^3 t e^{i V t_{34}/m} \left( G_{2/m}^+(t_{12}) +
G_{2/m}^-(t_{12})\right) \left[ G_{2/m}^+(t_{34}) (K_{1/m}^{-+} - K_{1/m}^{--}) +
G_{2/m}^-(t_{34})(K_{1/m}^{+-} - K_{1/m}^{++})\right].
\end{equation}
The cross correlation is given by
\begin{equation}  \Delta S_{b2,b1} =   { v_1^2 v_2^2\over m^2} \int
d^3 t e^{i V t_{34}/m} \left( G_{2/m}^+(t_{12}) -
G_{2/m}^-(t_{12})\right) \left[ G_{2/m}^+(t_{34}) K_{1/m}^{-+}  -
G_{2/m}^-(t_{34})K_{1/m}^{+-} \right].
\end{equation}

\subsection{Current for $m=2$}

In this section provide details of the calculation relating (4.17) to
(4.18) in the evaluation of
\begin{equation}
A(x_1,x_2) = \langle O_{2v}^+ (x_1)  O_{2v}^-(x_2) -O_{2v}^-(x_1) O_{2v}^+(x_2)\rangle.
\end{equation}
 The procedure is quite similar to that of appendix A1.  We
begin by rewriting (4.17) as
\begin{equation}
A(x_1,x_2)  = {1\over 2} v_1^2  \sum_{\{\sigma_k\}}
(\sigma_2-\sigma_1) \sigma_3 \sigma_4 \int dt_3 dt_4
 e^{i V t_{34}/m } \Pi(\{\sigma_k,t_k,x_k\})
 \end{equation}
 with
\begin{equation}
\Pi(\{\sigma_k,t_k,x_k\}) = \left\langle T_C\left[ O^+_{2v}(\sigma_1 0,x_1)
 O^-_{2v}(\sigma_2 0,x_2) O^-_{1v}(\sigma_3 t_3)  O^+_{1v}(\sigma_4 t_4)
 \right]\right\rangle_0.
\end{equation}
The correlation function has the same structure as (A5)
\begin{equation}
\Pi(\{\sigma_k,t_k,x_k\}) = G^{\sigma_{12}}_1(x_{12})
 G^{\sigma_{34}}_1(t_{34}) K_{1/2}^{\sigma_3 \sigma_4}
 \end{equation}
 with $G^\pm(x)$ and $K_{1/2}^{\sigma_3 \sigma_4}$ given in (A12,13)
 with $t_{1,2}$ replaced by $x_{1,2}$.  Summing on the Keldysh indices we find,
 \begin{equation}
A(x_1,x_2)  =  v_1^2 \int dt_3 dt_4 e^{i V t_{34}/m}
\left( G_1^+(x_{12}) -
G_1^-(x_{12})\right) \left[ G_1^+(x_{34}) (K_{1/2}^{-+} - K_{1/2}^{--}) +
G_1^-(t_{34})(K_{1/2}^{+-} - K_{1/2}^{++})\right].
\end{equation}
The first term in the integrand can be interpreted as the zeroth order expectation
value,
\begin{equation}
A^0(x_1,x_2) = \left\langle O_{2v}^+ (x_1)  O_{2v}^-(x_2) -O_{2v}^-(x_1)
O_{2v}^+(x_2)\right\rangle_0
= G_1^+(x_{12}) - G_1^-(x_{12}).
\end{equation}

\subsection{Noise for $m=2$}

Calculation of the expectation values $A_{1,2}(\{x_k\})$ in equations (4.25) and (4.26) of section IVB
can be done in the same manner as the previous section.  Again, the
expectation value can be factored into a zeroth order expectation value
times an integral.  We find
\begin{equation}
A_{1,2}(\{x_k\}) = v_1^2 A_{1,2}^0(\{x_k\}) \int dt_5 dt_6 e^{iV t_{34}/2}
\left[G^+(t_{56})(K^{-+}-K^{--}) +
G^-(t_{56})(K^{+-}-K^{++})\right]
\end{equation}
where $G^\pm(t)$ is the same as (A12)  and
\begin{equation}
K^{\sigma_5 \sigma_6} = \left({
\sin \pi T(\eta + i \sigma_5 z_{15})\sin \pi T(\eta + i \sigma_5 z_{35})
\sin \pi T(\eta + i \sigma_6 z_{26})\sin \pi T(\eta + i \sigma_6z_{46})
\over{
\sin \pi T(\eta + i \sigma_5 z_{25})\sin \pi T(\eta + i \sigma_5 z_{45})
\sin \pi T(\eta + i \sigma_6 z_{16})\sin \pi T(\eta + i \sigma_6 z_{36})
}}\right)^{1/2},
\end{equation}
where we use the notation $z_{ij} = x_i-t_j$.
The zeroth order expectation values can be evaluated using Wick's
theorem for the fermionic operators $O_{2v}^\pm$,
\begin{equation}
A_1^0(\{x_k\}) = \left(G^+(x_{12})-G^-(x_{12})\right)\left(G^+(x_{34})-G^-(x_{34})\right)
+ 4 G^+(x_{14}) G^+(x_{23})
\end{equation}
\begin{equation}
A_2^0(\{x_k\}) = G^+(x_{14}) G^-(x_{23})  - G^+(x_{12}) G^+(x_{34})
+ G^-(x_{14}) G^+(x_{23})-G^-(x_{12}) G^-(x_{34}).
\end{equation}

\section{Evaluation of Integrals}

\subsection{Small $t_2$ Perturbation Theory}
In this section we simplify the integrals (3.10) and (3.11).  One of the integrals can
be easily done because
\begin{equation}
K_1^{-+} - K_1^{--} = (K_1^{+-} - K_1^{++})^* = -(2/T) \delta(t_{14}){ \sin \pi T i t_{34}\ \
\sin \pi T i t_{12} \over
{\sin \pi T(\eta - i t_{23})}}
\end{equation}
This allows us to write the current ($I= C_-$) and noise ($S=C_+$) as
\begin{equation}
C_\pm  = -v_1^2 t_2^2 \int dt_2 dt_3 e^{iV t_3/m}\left(G^+_{2m-1}(t_2) \pm G^-_{2m-1}(t_2)\right)
\left(G^+_{2/m-1}(t_3) G^-_1(t_{23}) + G^-_{2/m-1}(t_3) G^+_1(t_{23})\right).
\end{equation}
Defining $u = \pi T t_2 \pm i \pi/2$ for the terms involving $G^\pm_{2m-1}(t_2)$
and $v = \pi T t_3\pm i \pi/2$ for the terms involving $G^\pm_{2/m-1}(t_3)$,
the terms in the integral can be combined and written in the scaling form
$C_\pm(V,T) = v_1^2 t_2^2 T^{2m+2/m-3} \tilde C_\pm(V/T)$ with
\begin{equation}
\tilde C_{t,m}(X)
= {2^{2-2m-2/m}\sinh(X/2m)\over\pi^2}\int_{-\infty}^\infty du dv
{ e^{i X v/m\pi}\over{\cosh^{2m-1}  u \cosh^{2/m-1} v }}\left({1\over{\sin(\eta+i(u-v))}} \mp
{1\over{\sin(\eta -i(u-v))}} \right)
\end{equation}
The integrals for $\tilde I_{t,m}(V/T)$ can be evaluated because the factor in parentheses is a $\delta$
function.  The result is given in (3.16).  The integral for $\tilde S_{t,m}(V/T)$ has the form
\begin{equation}
\tilde S_{t,m}(X) = {2^{3-2m-2/m}\sinh(X/2m)\over{\pi^2}}\int_{-\infty}^\infty du dv
{\sin X v/m\pi \over{\cosh^{2m-1}  u \cosh^{2/m-1} v  \sinh(u-v)}}.
\end{equation}
This is evaluated numerically in section V.

In special cases the above results simplify.  For $m=1$ we find
$\tilde I_{t,1}(X) = X/2\pi$
and
$\tilde S_{t,1}(X) = (X \coth X/2 - 2)/2\pi$.
Thus,
\begin{equation}
\tilde {\cal Q}_1(X,\infty) = \coth(X/2) - 2/X.
\end{equation}
These results are the same as those you get for non interacting electrons\cite{lesovik}.
For $m=2$ we find $\tilde I_{t,2}(X) = (1/32\pi^2)(X^2+ 4\pi^2)\tanh(X/4)$ and
$\tilde S_{t,2}(X) = (1/32\pi^2)(X^2+ 4\pi^2)\tanh^2(X/4)$.  Thus,
\begin{equation}
\tilde {\cal Q}_2(X,\infty) = \tanh(X/4).
\end{equation}

\subsection{Small $v_2$ Perturbation Theory}

In this section we evaluate the integrals for the correction to the current and noise at order $v_2^2$.
Since the purpose of this calculation is to establish the divergence of the perturbation
theory for $T\rightarrow 0$ with fixed $V$, we will focus on the limit $V \gg T$.

We begin with Eq. A26 for the current.  For $V/T \rightarrow\infty$ and $m>1$
the integral over $t_{34}$ is dominated by the region with $t_{34} \ll t_{31},t_{32}$, where
$K^{\sigma_3\sigma_4}_{1/m}$ is independent of $t_{34}$.  The integral over $t_3$ can then be
evaluated (with $t_4=0$) giving,
\begin{equation}
I_{b2} = v_1^2 v_2^2{(V/(2\pi m))^{2/m-1}\over{\Gamma(2/m)}}
 \int dt_1 dt_2 (G^+_{2/m}(t_{12}) - G^-_{2/m}(t_{12})) (K^{-+}_{1/m} - K^{--}_{1/m})
\end{equation}
Using the fact that $G^\pm_{2/m}(t_{12}) = e^{\mp i s_{12} \pi/m}(T/2\sin \pi T|t_{12}|)^{2/m}$
and $K^{\sigma_3\sigma_4}_{1/m} = e^{i (\pi/2m)(\sigma_3-\sigma_4)(s_{10}-s_{20})}$ (for $t_3=t_4=0$)
we then obtain
\begin{equation}
I_{b2} = a_m v_1^2 v_2^2 V^{2/m-1} T^{2/m-2}
\end{equation}
with
\begin{equation}
a_m ={1 \over{(2\pi m)^{2/m}}}{\Gamma(1/m)^2\over{
\Gamma(2/m)^2}} \sin(2\pi/m)
\end{equation}
Note that $a_2=0$.  The $v_2^2$ correction to the current vanishes for $V \gg T$ for $m=2$.

A similar calculation for the noise gives
\begin{equation}
\Delta S = b_m v_1^2 v_2^2 V^{2/m-1} T^{2/m-2}.
\end{equation}
$b_m$ has contributions from the four terms in (A23),
$b_m = b_{b2,b2, m} - 2 b_{b1,b2, m} + 2 b_{b2, 2 {\rm in}, m} - 2 b_{b2, 3 {\rm in}, m}$.
The first two terms can be evaluated by applying the analysis in eq (B7) to equations (A28) and (A29).
We find $b_{b2,b2,m} = b_{b1,b2,m} = a_m/m$, with $a_m$ given above.  The third and
fourth terms are evaluated by differentiating with respect to $V_2$ and $V_3$.
The dominant contribution for $V \gg T$ is due the term where the differentiation pulls down a factor
of $i t_{12}/m$.  Following the above analysis we then find
\begin{equation}
b_m =  {4\over{m(2\pi m)^{2/m}}} {\Gamma(1/m)^2\over{\Gamma(2/m)^2}} {\sin^3(\pi/m)\over{\cos(\pi/m)}}
\left({2\over\pi^2}\psi'(1/m) - 1\right)-{a_m\over m} .
\end{equation}
where $\psi'(x)$ is the derivative of the digamma function.
The coefficients can be evaluated for $m=2,3$ to be
$a_2 = 0$, $b_2 = 14 \zeta(3)/\pi^3 = 0.5428$.
$a_3 = 0.4786$, $b_3 = 0.8414$.
The effective charge then has the expansion
\begin{equation}
\tilde {\cal Q}_m(v_2/T_{1-1/m},V/T\rightarrow\infty) = {1\over m} + c_m{v_2^2\over T^{2-2/m}}
\end{equation}
with
\begin{equation}
c_m = {2\over{m\pi}} {\sin^3(\pi/m)\over{\cos(\pi/m)}}{\Gamma(1/m)^2\over{\Gamma(2/m)}}
\left({2\over\pi^2}\psi'(1/m)-1\right).
\end{equation}
Then $c_1=0$, $c_2 = 28 \zeta(3)/\pi^3 = 1.0855$ and $c_3 = 1.5279$.

\subsection{Exact Current $m=2$}

In this section we evaluate the integrals for the exact calculation of the
current for $m=2$ described in section IVB.  Combining (4.15) and (4.18) we find
\begin{equation}
I =
{ v_1^2 \over 2L} \int dx_1 dx_2   d_1(x_{12})
   \left( G_1^+(x_{12}) -
G_1^-(x_{12})\right)\int dt_3 dt_4 e^{i V t_{34}/2} \left[ G_1^+(t_{34})\left(K_{1/2}^{-+} -K_{1/2}^{--}\right) +
G_1^-(t_{34}) \left(K_{1/2}^{+-} -  K_{1/2}^{++}\right)\right].
\end{equation}
Using the fact that $G_1^+(t_{34}-i\pi T) = G_1^-(t_{34})$ and similar identities for $K$, we found
it convenient to rewrite the integral over $t_{3,4}$ as
\begin{equation}
\tanh{V\over 4T} \int dt_3 dt_4
e^{i V t_{34}/2} \left[ G_1^+(t_{34})\left(K_{1/2}^{-+} -K_{1/2}^{--}\right) -
G_1^-(t_{34}) \left(K_{1/2}^{+-} -  K_{1/2}^{++}\right)\right].
\end{equation}
The integration is then simplified using $G_1^+(t_{34}) + G_1^-(t_{34}) = \pi T \delta(t_{34})$.
The term involving $\delta(t_{34})$ does not contribute because
$K_{1/2}^{-+} -K_{1/2}^{--} - K_{1/2}^{+-} +  K_{1/2}^{++} = 0$ for $t_3=t_4$.
Then the integral over $t_{3,4}$ is then
\begin{equation}
2 \tanh{V\over 4T} \int dt_3 dt_4
e^{-i V t_{34}/2}   \left( G_1^+(t_{34}) -
G_1^-(t_{34})\right) \rho_{12}(t_3,t_4)
\end{equation}
with
\begin{equation}
\rho_{12}(t_3,t_4)  ={1\over 4} \left(K_{1/2}^{-+} + K_{1/2}^{+-} - K_{1/2}^{--} - K_{1/2}^{++}\right).
\end{equation}
This integral can be further simplified by symmetrizing the integrand with respect to permutations of
$t_3$ and $t_4$ and permutations of $x_1$ and $x_2$, and then restricting the
integration region to be $x_1>x_2$ and $t_3>t_4$  We then set $t_4=0$ to cancel the $L$.
Using a trigonometric identity it can be shown that
\begin{equation}
\rho_{12}(t_3,0) - \rho_{21}(t_3,0) = {\sinh \pi T t_3 \sinh\pi T x_{12}
\over {\sqrt{\sinh\pi T(x_1-t_3) \sinh\pi T(t_3-x_2) \sinh \pi T(x_1) \sinh\pi T(- x_2)}}}
\end{equation}
when $x_2<0<t_3<x_1$ and $0$ otherwise.  We then find
\begin{equation}
I =  2  v_1^2T^2 \tanh{V\over 4T} \int_0^\infty dx_1 \int_{-\infty}^0 dx_2 \int_0^{x_1} dt_3
{\left(\delta(x_{12}) - v_2^2 e^{-2 v_2^2|x_{12}|}\right)\cos(Vt_3/2)
\over\sqrt{\sinh\pi T(x_1-t_3) \sinh\pi T(t_3-x_2) \sinh \pi Tx_1 \sinh\pi T(- x_2)}}
\end{equation}

The two terms in parentheses in (B19) can be interpreted
as the incident and backscattered currents for the second junction,
$I = I_{\rm in} - I_{b2}$.
The $\delta$ function term can be evaluated using a concrete regularization of the
$\delta$ function,
$
\delta(x_{12}) = \lim_{Z\rightarrow\infty} Z \exp(-2 Z |x_{12}|).
$
This gives
\begin{equation}
I_{\rm in} = {v_1^2\over2} \tanh{V\over 4T},
\end{equation}
in agreement with the current calculated for $v_2=0$ in section III for $m=2$.

For the second term we define new variables, $y_1 = \pi T(  x_1-t_3) $,
$y_2 = -\pi T   x_2$, $u = \pi T t_3$.
The backscattered current can then be written in the form,
\begin{equation}
I_{b2} =   {2v_1^2 v_2^2\over{\pi^3 T}} \tanh{V\over 4T} \int_0^\infty du dy_1 dy_2
{e^{ -2  (u+y_1+y_2)v_2^2/\pi T}
\cos( V u/2\pi T)
\over{\sqrt{\sinh   y_1 \sinh   (u+y_1) \sinh  y_2 \sinh (u+y_2) }}}
\end{equation}.

Combining (B20) and (B21) the final result can be cast
in the scaling form  $I(V,T) = v_1^2 \tilde I(v_2/T^{1/2},V/T)$ with
\begin{equation}
\tilde I_2(v_2/T^{1/2},V/T) = {1\over 2} \tanh{V\over 4T}\left(
1 - {4 v_2^2\over {\pi^3 T}} \int_0^\infty du  \cos {u V\over 2\pi T}
\left[\int_0^\infty dy {e^{-(u+ 2 y) v_2^2/\pi T}\over\sqrt{\sinh y \sinh(y+u)}}\right]^2\right).
\end{equation}

The integrals can be evaluated in the limit of large and small $v_2$, with results quoted in section
IIIB.  In the limit of zero temperature, the integrals simplify.  Rescaling the integration
variables by $v_2^2/\pi T$ we may write the current in the form (4.20) with
\begin{equation}
\tilde I_2'(v_2/V^{1/2},\infty) = {1\over 2} \left(
1 - {4 \over {\pi ^2 }} \int_0^\infty du  \cos{Vu\over {2v_2^2}}
\left[\int_0^\infty dy {e^{-( u+ 2 y)}\over\sqrt{y (y+u)}}\right]^2\right).
\end{equation}
The integral over $y$ in the square brackets is a Bessel function, $K_0(u)$.
The remaining integral over $u$ then gives
\begin{equation}
\tilde I_2'(v_2/V^{1/2},\infty) = {1\over 2} \left[
1 - {2\over \pi} {\bf K}\left(-{V^2\over{16 v_2^4}}\right)\right].
\end{equation}
where ${\bf K}$ is the elliptic integral of the second kind.

\subsection{Exact Noise $m=2$}

Combining (4.24) and (A37) and using the transformations (B9-B12) the noise may be written,
\begin{equation}
\Delta S = {v_1^2\over 2L}\tanh{V\over 4T} \int d^4 x dt_5 dt_6 F(\{x_k\})
\left(G_1^+(t_{56})- G_1^-(t_{56})\right) \rho(\{x_k\},t_5,t_6) e^{i V t_{56}/2}
\end{equation}
where as $\rho(\{x_k\},t_5,t_6) = (K_{1/2}^{-+} + K_{1/2}^{+-} - K_{1/2}^{--} - K_{1/2}^{++})/4$ with
$K$ given in (A38) and
\begin{equation}
F(\{x_k\}) = d_1(x_{12}) d_1(x_{34}) A_1^0(\{x_k\})
- d_2(x_{13}) d_2(x_{24})  A_2^0(\{x_k\}).
\end{equation}
$A_{1,2}^0(\{x_k\})$ is given in (A39,40), and
$d_{1,2}(x)$ are in (4.13,4.14).
It is again useful to symmetrize the integrand with respect to permutations
of $x_1,x_2,x_3,x_4$ and permutations of $t_5$ and $t_6$.  $t_6$ is then set to zero,
and we define $y_k = \pi T x_k$, $u = \pi T t_5$.
After some lengthy algebra one finds,
\begin{equation}
\Delta S = {2 v_1^2\over\pi^3}\tanh{V\over 4T} \int_R d^4 y du  \Theta(\{y_k\},u)\sin{Vu\over {2\pi T}}
{D(\{y_k\}) M(y_1,y_3,u) M(y_2,y_4,u)\over{\sinh u \sinh y_{13}
\sinh \pi y_{24}}},
\end{equation}
where the integration region $R$ is $y_1>y_2>y_3>y_4$ and $u>0$.  In addition
\begin{equation}
D(\{y_k\}) =
\delta_{12} \delta_{34} +
{v_2^2\over{\pi T}}\left( \delta_{23} e^{-2 y_{14}v_2^2/\pi T}-
\delta_{12} e^{-2 y_{34}v_2^2/\pi T} - \delta_{34}
e^{-2  y_{12}v_2^2/\pi T}\right) + {v_2^4\over{\pi^2 T^2}}  e^{-2 (y_{12}+y_{34})v_2^2/\pi T},
\end{equation}
\begin{equation}
M(y_i,y_j,u) = { \sinh y_i \sinh (y_j-u) + \sinh  y_j \sinh (y_i-u) \over
\sqrt{| \sinh y_i \sinh y_j \sinh  (y_j-u)
\sinh  (y_i-u)|}},
\end{equation}
and
\begin{equation}
\Theta(\{y_k\},u) = \left\{
\begin{array}{cl}
1 &{\rm for}\quad y_1>u>y_2>y_3>0>y_4\\
-1 &{\rm for}\quad y_1>y_2>y_3>u>0>y_4\\
-1 &{\rm for}\quad y_1>u>0>y_2>y_3>y_4\\
0 &{\rm otherwise}.
\end{array}\right.
\end{equation}
We have evaluated these integrals numerically to obtain the scaling function
$\tilde S(v_2/T^{1/2},V/T)$.  The results were discussed in section V.

In the
limit of zero temperature  it is possible to obtain an analytic solution.
Due to the complexity of the integral and to explain a subtlety in dealing with the
$\delta$ functions we  divide the result into three contributions by writing
$I = I_{\rm in} - I_{b2}$, where $I_{\rm in}$ is the current
incident on the 2nd junction and $I_{b2}$ is the current
backscattered by the second junction.  The noise is then a sum of
three terms,
\begin{equation}
\Delta S = \Delta S_{\rm in,in}  - 2 \Delta S_{\rm in,b2}+ \Delta S_{\rm b2,b2}.
\end{equation}
These three terms arise from the three terms in $D(\{x_k\})$.

The term with two $\delta$ functions gives the noise incident on the second junction.
It can be evaluated using the regularization $
\delta_{ij} \equiv \delta(y_{ij}) = \lim_{Z\rightarrow\infty} Z \exp(-2 Z |y_{ij}|)
$.  We find
\begin{equation}
\Delta S_{\rm in,in} = {v_1^2\over 4} \tanh^2({V\over 4T}),
\end{equation}
in agreement with the result for $v_2=0$ discussed in section
IIIA.  Thus at zero temperature $\Delta S_{\rm in,in}(T=0) = v_1^2/4$.

 The terms with a single delta function describe the cross correlations
between $I_{b2}$ and $I_{\rm in}$.  Again using the regularized $\delta$
 function two of the integrals in (B23) can be evaluated
analytically.  At finite temperature the
remaining three integrals must be evaluated numerically.  At
 zero temperature, however, the cross correlation is simply related
to the backscattered current computed in section B3,
\begin{equation}
\Delta S_{b2, \rm in}(T=0) = {1\over 2} I_{b2} = {1\over {2\pi}}{\bf K}
\left({V^2\over{16v_2^4}}\right).
\end{equation}

The final term in (B24) describes the backscattered noise.
At zero temperature we may write
$\Delta S_{b2,b2} = v_1^2 \tilde S_{b2,b2}'(v_2/V^{1/2})$ with
\begin{equation}
\tilde S_{b2,b2}'(X) = {2\over{\pi^3}}\int_R d^4y du \Theta(\{y_k\},u)
\sin\left( {u\over 2X^2}\right) {
 e^{-2(y_{12}+y_{34})} \over{u y_{13}y_{24}}}
{(y_1(y_3-u) + y_3(y_1-u))(y_2(y_4-u) + y_4(y_2-u))\over{
\left|y_1(y_1-u)y_2(y_2-u)y_3(y_3-u)y_4(y_4-u)\right|^{1/2}}}
\end{equation}
While we have been unable to evaluate this integral analytically, we computed it numerically
and found that $\tilde S_{b2,b2}'(X) = 1/4$ independent of $X$.  We checked this
result analytically in the limits of large and small $X$.   We
thus conclude that the noise, when written in the scaling form is
\begin{equation}
\tilde S'(v_2/V^{1/2},\infty) = {1\over 2}\left[ 1 - {2\over \pi}
{\bf K}\left({V^2\over{16v_2^4}}\right)\right].
\end{equation}
This is exactly the same as the transmitted current (B24), so the shot noise is due
to electrons, even in the weak backscattering limit.

The limiting behavior of $\tilde S(v_2/T^{1/2},V/T)$ for $v_2=0$ is given by (B28), in
agreement with the results of section IIIA.  For large $v_2$ we may write
$D(\{y_k\}) = (\pi^2 T^2/4 v_2^4) \delta'(y_{12}) \delta'(y_{34})$, where
$\delta'(y) = \lim_{Z\rightarrow\infty} 2Z^2 {\rm sign}(y) \exp(-2 Z y)$.  This leads
to integrals identical to those of section IIIB.

\end{appendix}

\end{document}